\documentclass[a4paper,aps,prd,showpacs,preprintnumbers,twocolumn,nofootinbib,superscriptaddress]{revtex4-1}

\usepackage[english]{babel}

\usepackage[utf8x]{inputenc}
\usepackage[T1]{fontenc}
\usepackage{ucs}
\usepackage{microtype}
\usepackage{newtxtext,newtxmath}
\usepackage{color}

\usepackage{url}
\usepackage{hyperref}

\usepackage{xspace}
\usepackage{lastpage}
\usepackage{stmaryrd}

\usepackage{graphicx}
\usepackage{empheq}
\usepackage{ifthen}
\usepackage{tensor}
\usepackage{paralist}

\usepackage{amsmath}
\usepackage{amsfonts}
\usepackage{amssymb}
\usepackage{mathrsfs}
\usepackage{extarrows}
\usepackage{verbatim}
\usepackage{upgreek}
\usepackage{wasysym}

\usepackage{natbib}
\newcommand\eps{\ensuremath{\varepsilon}}

\newcommand{\cst}{\mathrm{cst}}

\newcommand\define{\equiv}

\newcommand\vect[1]{\boldsymbol{#1}}
\newcommand{\unit}[1]{\hat{\vect{#1}}}

\newcommand\ex[1]{\mathrm{e}^{#1}}
\renewcommand\i{\ensuremath{\mathrm{i}}}

\newcommand\e[1]{_{\text{#1}}}
\newcommand\h[1]{^{\text{#1}}}
\newcommand\U[1]{\:\mathrm{#1}}

\newcommand{\dd}{\mathrm{d}}

\newcommand{\pd}[3][]{\frac{\partial^{#1} #2}{\partial {#3}^{#1}}}
\newcommand{\ddf}[3][]{\frac{\dd^{#1} #2}{\dd {#3}^{#1}}}

\newcommand{\delimiters}[4][]{
\ifthenelse{ \equal{#1}{1} }{  #2 #3 #4  }
					{ \ifthenelse{\equal{#1}{2}}{ \big#2 #3 \big#4 }
						{ \ifthenelse{\equal{#1}{3}}{ \Big#2 #3 \Big#4 }
							{ \ifthenelse{\equal{#1}{4}}{ \bigg#2 #3 \bigg#4 }
								{ \ifthenelse{\equal{#1}{5}}{ \Bigg#2 #3 \Bigg#4 }
									{ \left#2 #3 \right#4 }
								}
							}
						}
					}
													}

\newcommand{\pa}[2][]{\delimiters[#1]{(}{#2}{)}}
\newcommand{\pac}[2][]{\delimiters[#1]{[}{#2}{]}}
\newcommand{\paac}[2][]{\delimiters[#1]{\{}{#2}{\}}}
\newcommand{\abs}[2][]{\delimiters[#1]{|}{#2}{|}}
\newcommand{\ev}[2][]{\delimiters[#1]{\langle}{#2}{\rangle}}

\newcommand{\apjs}{ApJS}
\newcommand{\apjl}{Astrophys. J. Lett.}
\newcommand{\aap}{A{\&}A}

\newcommand{\mnras}{MNRAS}

\renewcommand{\prl}{Phys. Rev. Lett.}

\newcommand{\Pei}{P93\xspace}

\begin{document}

\title{On simple analytic models of microlensing amplification statistics}

\author{Pierre Fleury}
\email{pierre.fleury@uam.es}
\affiliation{Instituto de F\'isica Te\'orica UAM-CSIC,
Universidad Auton\'oma de Madrid,\\
Cantoblanco, 28049 Madrid, Spain}
\affiliation{D\'{e}partment de Physique Th\'{e}orique, Universit\'{e} de Gen\`{e}ve,\\
24 quai Ernest-Ansermet, 1211 Gen\`{e}ve 4, Switzerland}

\author{Juan Garc\'ia-Bellido}
\email{juan.garciabellido@uam.es}
\affiliation{Instituto de F\'isica Te\'orica UAM-CSIC,
Universidad Auton\'oma de Madrid,\\
Cantoblanco, 28049 Madrid, Spain}

\begin{abstract}
Gravitational microlensing is a key probe of the nature of dark matter and its distribution on the smallest scales. For many practical purposes, confronting theory to observation requires to model the probability that a light source is highly amplified by many-lens systems. This article reviews four simple analytic models of the amplification probability distribution, based on different approximations: (i) the strongest-lens model; (ii) the multiplicative model, where the total amplification is assumed to be the product of all the lenses' individual amplifications; (iii) a hybrid version of the previous two; and (iv) an empirical fitting function. In particular, a new derivation of the multiplicative amplification distribution is proposed, thereby correcting errors in the literature. Finally, the accuracy of these models is tested against ray-shooting simulations. They all produce excellent results as long as lenses are light and rare (low optical depth); however, for larger optical depths, none of them succeeds in capturing the relevant features of the amplification distribution. This conclusion emphasizes the crucial role of lens-lens coupling at large optical depths.
\end{abstract}

\date{\today}
%\preprint{}
\maketitle

%%%%%%%%%%%%
\section{Introduction}
%%%%%%%%%%%%

When, in 1936, Einstein was convinced by Mandl to publish the outcome of a short calculation about the \emph{lens-like action of a star by the deviation of light in the gravitational field}~\cite{1936Sci....84..506E}, he could not imagine the astronomical potential of his finding. That research note already proposed that the apparent brightness of a light source can be highly amplified by compact forms of matter on the line of sight. Somehow excavated 27 years later by Liebes~\cite{1964PhRv..133..835L}, this topic mostly concerned, until the mid-1980s, the effect of stars in galaxies acting as lenses for distant quasars~\cite{1979Natur.282..561C, 1981ApJ...243..140G, 1981ApJ...244..756Y, 1984A&A...132..168C}. The term \emph{microlensing} seems to have been introduced in 1986 by Paczy\'{n}ski~\cite{1986ApJ...301..503P}, who also proposed to use it as a probe of MAssive Compact Halo Objects (MACHO) in our galaxy~\cite{1986ApJ...304....1P}. This idea led to several surveys in the 1990s-2000s: the MACHO experiment~\cite{Alcock:1995zx}; the Exp\'{e}rience pour la Recherche d'Objets Sombres (EROS)~\cite{2003A&A...400..951A}; and the still ongoing Optical Gravitational Lensing Experiment (OGLE)~\cite{2015ApJS..216...12W}.

While focus was later displaced towards the detection of exoplanets~\cite{1991ApJ...374L..37M, Bond_2004, Ranc}, the concept of microlensing as a probe of the nature and distribution of dark matter never ceased to be enriched with new ideas, from multiply-imaged quasars~\cite{2017ApJ...836L..18M} to supernova lensing~\cite{Seljak:1999tm, Metcalf:1999qb, Metcalf:2006ms}. More recently, interest in that matter was revived by the observation of \emph{Icarus}~\cite{Kelly:2017fps}, a single star visible through cosmological distances thanks to a huge gravitational amplification, on the order of $10^3$. The very possibility of such an event was attributed to the disruption of a strong lens' caustic by its own substructure~\cite{Oguri:2017ock, Diego:2017drh}, thereby opening a new branch in gravitational-lensing science~\cite{2017ApJ...850...49V, Diego:2018fzr, 2018ApJ...867...24D, 2019ApJ...880...58K, 2019arXiv190801773D}.

In many concrete microlensing problems, a central observable is the amplification probability distribution function (PDF), $p(A)$. From the theoretical side, the difficulty consists in accurately relating this PDF to the matter distribution producing the amplifications. A significant research endeavor~\cite{1986ApJ...306....2K, 1987PhRvL..59.2814D} was conducted in that direction during the 1980s-1990s, presumably with the hope to explain the variability of quasars with microlensing~\cite{1987A&A...171...49S, 1994A&A...288....1S, 1994A&A...288...19S, 2010MNRAS.405.1940H, 2014MNRAS.441.1708G}. In that context, it was first understood in Ref.~\cite{1982MNRAS.199..987P} that $p(A)$ generally displays a long algebraic tail, of the form $A^{-3}$, when microlensing is at work. Several attempts to further characterize the full PDF~\cite{1983ApJ...267..488V,1986MNRAS.223..113P, 1991A&A...251..393M, 1993ApJ...403....7P} then led to a better understanding of the nonlinear coupling between lenses, when these are very numerous~\cite{1984JApA....5..235N, 1986ApJ...310..568B, 1987ApJ...319....9S, 1990ApJ...357...23L, 1992ApJ...389...63M, 1997ApJ...489..508K, 1997ApJ...489..522L}. Theoretical works were also guided by and tested with numerical simulations~\cite{1986A&A...166...36K, 1992ApJ...386...19W, 1991ApJ...374...83R, 1995MNRAS.276..103L}. Finally, Refs.~\cite{2009JMP....50g2503P, 2009JMP....50l2501P, 2010GReGr..42.2011P} must be mentioned for a number of results on the mathematics of gravitational lensing.

On the shoulders of giants, the goal of the present article is rather modest. The aforementioned works on amplification statistics have led to several modeling techniques, the simplest ones being currently used to set or forecast constraints on the nature of dark matter~\cite{Zumalacarregui:2017qqd, Garcia-Bellido:2017imq, Diego:2018fzr}. However, it seems that the performance of these simple models for $p(A)$ has never been properly assessed. We propose here to fill this gap. Such a comparative analysis will also be the occasion to clarify and correct some theoretical points of the existing literature.

Specifically, four models will be reviewed throughout this article. In Sec.~\ref{sec:individual}, we introduce microlensing fundamentals, and consider the \emph{strongest-lens} model, where the amplification due to a set of lenses is strictly due to the strongest one. In Sec.~\ref{sec:multiplicative} we consider the \emph{multiplicative} approach, where the total amplification, due to many lenses, is assumed to be the product of all their individual amplifications. We use this opportunity to correct the corresponding derivation of $p(A)$ with respect to earlier works. We also propose a \emph{hybrid} model between the strongest-lens and the multiplicative models. Finite-source corrections to the previous calculations are considered in Sec.~\ref{sec:extended_sources}. Finally, in Sec.~\ref{sec:numerics}, the above two approaches are compared with an \emph{empirical} expression for $p(A)$ and confronted to numerical ray-shooting simulations. We conclude in Sec.~\ref{sec:conclusion}.

We adopt units in which the speed of light is unity, $c=1$. Bold symbols, such as~$\vect{\theta}$, stand for two-dimensional vectors. A hatted vector~$\unit{\theta}$ denotes the unit counterpart to the nonhatted one, $\unit{\theta}=\vect{\theta}/\theta$, where $\theta\define ||\vect{\theta}||$ is the Euclidean norm of $\vect{\theta}$.

%%%%%%%%%%%%%%%%%%%%%%%
\section{Microlensing by individual lenses}
\label{sec:individual}
%%%%%%%%%%%%%%%%%%%%%%%

\subsection{Amplification by a point lens}
\label{subsec:point_lens}

\begin{figure}[h!]
\centering
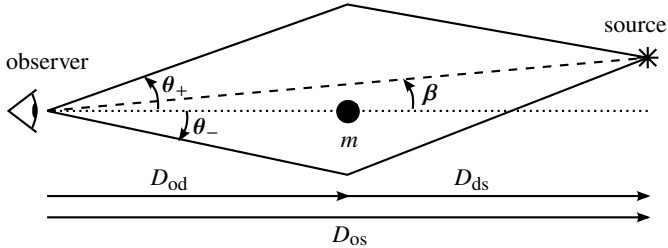
\caption{Geometry of the lens equation. A source with unlensed position $\vect{\beta}$ has two images at $\vect{\theta}_\pm$.}
\label{fig:lens_equation}
\end{figure}

Consider an infinitesimal light source lensed by a point mass. Throughout this article, we will work in the weak gravitational field and geometric optics regimes, assuming the small-angle and flat-sky approximations---see Ref.~\cite{Fleury:2018cro} for a comparative discussion of these assumptions. Let the lens be at the origin of a celestial coordinate system, and call~$\vect{\beta}$ the unlensed position of the source (see Fig.~\ref{fig:lens_equation}). The image position~$\vect{\theta}$ then satisfies the lens equation~\cite{1992grle.book.....S}
\begin{equation}\label{eq:lens_eq_1PL}
\vect{\beta} = \vect{\theta} - \frac{\eps^2}{\vect{\theta}} \ ,
\end{equation}
where $1/\vect{\theta}\define \unit{\theta}/\theta$, while
\begin{equation}\label{eq:Einstein_radius_def}
\eps^2 \define \frac{4 Gm D\e{ds}}{D\e{d}D\e{s}}
\end{equation}
denotes the Einstein radius of the lens. In Eq.~\eqref{eq:Einstein_radius_def}, $m$ denotes the mass of the lens; $D\e{d}, D\e{s}$ are respectively the angular-diameter distance of the lens (deflector), source, as seen from the observer; and $D\e{ds}$ is the distance to the source as seen from the lens. The precise expression of these distances depend on a choice of background, i.e., a fiducial no-lensing situation.

The lens equation~\eqref{eq:lens_eq_1PL} has two solutions~$\vect{\theta}_\pm=\theta_\pm \unit{\beta}$
\begin{equation}
\theta_\pm = \frac{1}{2} \pa{ \beta \pm \sqrt{\beta^2+4\eps^2} } ,
\end{equation}
which are the positions of the two images of the source at $\vect{\beta}$. The luminous intensity of each image reads~$I_\pm=A_\pm I_0$, where $I_0$ is the unlensed intensity, i.e. the source's apparent luminosity in the absence of lensing. Due to surface-brightness conservation, the individual inverse amplifications (or magnifications) read
\begin{equation}\label{eq:A_pm_u}
A_\pm^{-1}
= \abs{ \det \pd{\vect{\beta}}{\vect{\theta}} } (\vect{\theta}_\pm)
= \abs{ 1 - \frac{\eps^4}{\theta_\pm^4} }
= \abs{ \frac{1}{2} \pm \frac{1}{2} \frac{u^2+2}{u\sqrt{u^2+4}} }^{-1} \ ,
\end{equation}
with $u\define \beta/\eps$.

If the two images are not resolved by the telescope, i.e. if $|\theta_+-\theta_-|$ is smaller than the telescope's resolution, they are called microimages. Lensing then only manifests through the apparent amplification of the luminosity of the macro-image, $I=I_++I_-$. This is a microlensing event, and the total amplification reads
\begin{equation}\label{eq:A_u}
A(u) = A_++A_- = \frac{u^2 + 2}{u\sqrt{u^2+4}} \ .
\end{equation}
This relation can also be inverted to get
\begin{equation}\label{eq:u_A}
u^2 = \frac{2 A}{\sqrt{A^2-1}} - 2 \ .
\end{equation}

Finally, it is useful to rewrite $u$ as
\begin{equation}
u = \frac{b}{r\e{E}} \ ,
\end{equation}
where $b=D\e{d}\beta$ is the physical impact parameter of the unlensed light path, and
\begin{equation}\label{eq:Einstein_radius_physical}
r\e{E} \define D\e{d} \eps = \sqrt{4GM\mathcal{D}} \ ,
\qquad
\mathcal{D} \define \frac{D\e{d}D\e{ds}}{D\e{s}} \ .
\end{equation}
Thus, $u$ must be understood as a reduced impact parameter, i.e. expressed in units of the lens' cross-sectional radius. Indeed, it is customary to designate $\pi r\e{E}^2$, or $\pi\eps^2$, the cross section of the lens. This is because $\pi\eps^2$ is the area of the sky where a light source gets amplified by a factor $A>A(1)\approx 1.34$.

\subsection{From one to many: the strongest-lens approximation}
\label{subsec:strongest_lens}

The single-lens case is the only one which is analytically solvable. However, in many physically relevant situations, a given source may be affected by many lenses. These cases include microlensing by planetary systems (lenses are a star and its planets)~\cite{1991ApJ...374L..37M}; but also nearby supernovae or quasars observed through galaxies (lenses are stars and globular clusters)~\cite{1983ApJ...267..488V}; or distant sources observed through clusters of galaxies (lenses are galaxies). This question is especially relevant in a scenario where a significant fraction of dark matter would be made of compact objects, such as primordial black holes~\cite{Carr:1974nx, GarciaBellido:1996qt, 2016PhRvL.116t1301B, Carr:2016drx, Clesse:2017bsw}.

How to model the combined effect of many lenses? Specifically, we aim, here, to evaluate the probability density function (PDF) of the microlensing amplification, $p(A)$, due to all the lenses potentially located between the source and the observer. If the lenses are rare and not too massive, then the sum of their individual cross sections~$\pi\eps^2$ occupy a small fraction of the sky. Equivalently, $\eps\ll\Delta\theta$, where $\Delta\theta$ is the typical angular separation between two lenses. In that context, referred to as the \emph{low-optical-depth} regime, a typical source is mostly affected by a single lens: the one with smallest reduced impact parameter~$u$.

To the best of our knowledge, the explicit expression of $p(A)$ in the strongest-lens approximation was first derived by Peacock in 1986~\cite{1986MNRAS.223..113P}. Another derivation, originally due to Nottale, can can also be found in Ref.~\cite{1991A&A...251..393M}. In this section, we propose an alternative, step-by-step, calculation of $p(A)$, whose formalism will be useful in the remainder of the article.

\subsubsection{Setup and notation}
\label{subsubsec:setup}

\begin{figure}[h!]
\centering
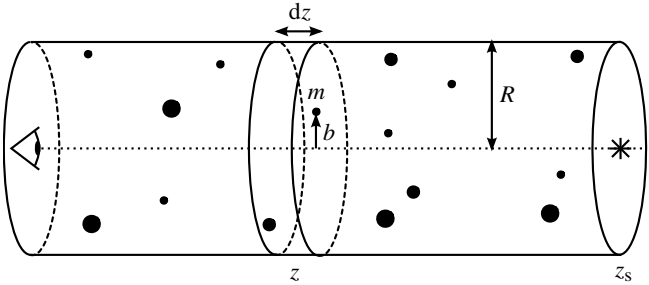
\caption{A tube of Universe with radius $R$ between the source and the observer, containing $N$ point lenses.}
\label{fig:tube}
\end{figure}

Consider, as depicted in Fig.~\ref{fig:tube}, a tube with physical radius~$R$ between the observer and the source, filled with a finite number $N$ of lenses. These lenses may have different masses, and their distribution may be inhomogeneous along the line of sight. Our only assumption is that, on each disk $z=\cst$, their spatial distribution is Poissonian.

For a single lens, the probability that its mass is in $[m, m+\dd m]$, its redshift in $[z, z+\dd z]$, and its physical impact parameter with respect to the line of sight is in $[b, b+\dd b]$, reads
\begin{equation}\label{eq:p_1_b}
p_1(m,z,b)\, \dd m\,\dd z\,\dd b = \frac{2b\,\dd b}{R^2} \, \Theta(R-b) \, p(m,z) \, \dd m\,\dd z \ .
\end{equation}
In Eq.~\eqref{eq:p_1_b}, the Heaviside function $\Theta$ ensures that the lens lies within the tube, while $p(m,z)$ is an arbitrary distribution of masses and redshifts, which may be correlated. The macroscopic number density of lenses, per unit area and redshift, is related to $p(m,z)$ as
\begin{equation}
\ddf{\Sigma}{z} \define \frac{\dd^3 N}{\dd^2 S \dd z} = \frac{N}{\pi R^2} \int_0^\infty p(m,z) \; \dd m \ ,
\end{equation}
while the mean surface density, obtained by stacking all the planes $z=\cst$, simply reads
\begin{equation}
\Sigma = \int_0^{z\e{s}} \ddf{\Sigma}{z}\;\dd z = \frac{N}{\pi R^2} \ .
\end{equation}
When, at the very end of the calculation, we will take $N, R\rightarrow \infty$, the surface density~$\Sigma$ will be kept constant, so that $N = \mathcal{O}(R^2)$.

\subsubsection{Amplification probability for one lens}

Let us first consider the case of a single lens in the entire tube, and determine the corresponding amplification PDF, $p_1(A)$. Because the $A$ only depends on the reduced impact parameter $u$, it is equivalent to determine $p_1(u)$. For that purpose, the first step consists in translating Eq.~\eqref{eq:p_1_b} in terms of $u=b/r\e{E}(m,z)$,
\begin{align}
p_1(m, z, u)
&= \frac{2 u \, r\e{E}^2(m,z)}{R^2} \, \Theta(R-u r\e{E}) \, p(m, z) \ .
\label{eq:p_1_u}
\end{align}
The PDF of $u$ only is then obtained by marginalizing over $m, z$. In that operation, the subtlety is how to handle the Heaviside function, which depends on both $z, m$ via $r\e{E}$. The simplest way consists in encoding it in an upper limit for the integration over $m$. We define $M(u,z,R)$ such that $R=u \, r\e{E}(M,z)$, which represents the mass above which the impact parameter~$b$ should be larger than $R$ to have the correct value of $u$. With this notation,
\begin{equation}\label{eq:p_1_with_M}
p_1(u)
= \frac{2 u}{R^2} \int_0^{z\e{s}} \dd z \int_0^{M(u,z,R)} \dd m \; p(m,z) \, r\e{E}^2(m,z) \ .
\end{equation}

Of course, in reality, the lens mass distribution does not extend to infinity; there exists some maximum lens mass $m\e{max}$ such that $p(m>m\e{max},z)=0$. Since the function $u\mapsto M(u,z,R)\propto u^{-2}$ is monotonically decreasing from $(0,\infty)$ to $(0,\infty)$, there exists a critical value $u\e{c}(R)$ such that $\forall z\in[0,z\e{s}] \quad M(u\leq u\e{c},z,R)\geq m\e{max}$. Its explicit expression is easily found to be
\begin{equation}\label{eq:u_c}
u\e{c} = \frac{R}{\sqrt{4 G m\e{max} \mathcal{D}\e{max}}} \ ,
\end{equation}
where $\mathcal{D}\e{max}$ is the maximum value of $\mathcal{D}$, which is typically reached when the lens lies midway between the observer and the source. Thus, for $u\leq u\e{c}$, we can substitute $M$ with $m\e{max}$ in Eq.~\eqref{eq:p_1_with_M} and get
\begin{equation}\label{eq:p_1_u_result}
\forall u \leq u\e{c} \qquad
p_1(u) = \frac{2u\ev[1]{r\e{E}^2}}{R^2} \ ,
\end{equation}
where we introduced
\begin{equation}
\ev[1]{r\e{E}^2} \define \int_0^{z\e{s}} \dd z \int_0^{m\e{max}} \dd m \; p(m,z) \, r\e{E}^2(m,z) \ .
\end{equation}
It is not necessary to explicitly determine the case $u\geq u\e{c}$ in order to proceed.

\subsubsection{The strongest out of $N$ lenses}

Let us now consider the case where $N$ lenses are in the tube, and let us determine the PDF of the strongest amplification. Again, since $A$ only depends on $u$, the strongest lens is the one with the smallest reduced impact parameter~$u$. Let us call $p\e{s}(u)$ the associated PDF (subscript ``s'' stands for ``strongest''). Assuming that the $N$ lenses are independent, we have
\begin{equation}
\label{eq:p_s_u_1}
p\e{s}(u) \, \dd u
= \sum_{i=1}^N
	p_i(u) \, \dd u \times \text{Prob}(u_{j\not= i}\geq u) \ ,
\end{equation}
where $p_i$ is the unconstrained PDF of $u$ for the $i$th lens. Since all the lenses are characterized by the same PDF~$p_1(u)$, Eq.~\eqref{eq:p_s_u_1} becomes
\begin{equation}\label{eq:p_s_u_N}
p\e{s}(u) = N p_1(u) \pac{ 1 - \int_0^u \dd u'\; p_1(u')}^{N-1} .
\end{equation}

\subsubsection{Infinite number of lenses}

We now take the limit of an infinite Universe containing an infinite number of lenses, i.e. $N, R \rightarrow\infty$, while keeping the surface density $\Sigma=N/(\pi R^2)$ fixed. In that limit, $u\e{c}\rightarrow \infty$, and hence, for any $u$,
\begin{equation}
N p_1(u) \rightarrow 2 \kappa u \ ,
\end{equation}
where we introduced the \emph{microlensing optical depth}
\begin{equation}
\label{eq:optical_depth}
\kappa
\define \Sigma \pi \ev[1]{r\e{E}^2}
= \int_0^{z\e{s}} \dd z \int_0^\infty \dd m\; \frac{\dd^2\Sigma}{\dd z \dd m} \, \pi r\e{E}^2(m,z) \ .
\end{equation}
Besides, the limit of the square-bracket term in \eqref{eq:p_s_u_N} yields an exponential, so that
\begin{equation}
p\e{s}(u)
= 2\kappa u \, \ex{-\kappa u^2}
= -\ddf{}{u} \pa{\ex{-\kappa u^2}} \ .
\end{equation}
This result is easily translated in terms of amplifications using Eq.~\eqref{eq:u_A}, and yields the PDF of the strongest amplification
\begin{empheq}[box=\fbox]{align}\label{eq:p_s_A}
p\e{s}(A)
&= \ddf{}{A} \exp\pac{ -2\kappa\pa{ \frac{A}{\sqrt{A^2-1}} - 1}} \\
&= \frac{2\kappa}{(A^2-1)^{3/2}} \, \exp\pac{ -2\kappa \pa{ \frac{A}{\sqrt{A^2-1}} - 1} } ,
\end{empheq}
in agreement with Refs.~\cite{1986MNRAS.223..113P, 1991A&A...251..393M}.

\subsection{Discussion: optical depth, Weinberg, and Peacock}

The most striking property of Eq.~\eqref{eq:p_s_A} is that it only depends on a single parameter: the optical depth~$\kappa$. This parameter notably controls the amplitude of the high-amplification tail of the distribution, $p\e{s}(A\gg 1)\sim 2\kappa/A^3$, which is a well-known behavior~\cite{1983ApJ...267..488V}. Alternatively, one could say that the entire PDF is controlled by its mean,
\begin{equation}
\ev{A}_s \define \int_1^\infty \dd A \; A\,p\e{s}(A) \ .
\end{equation}
With the change of variable~$\cosh x = A/\sqrt{A^2-1}$, one finds
\begin{align}
\ev{A}\e{s} &= 2\kappa \ex{2\kappa} \int_0^\infty \dd x \; \cosh x \, \ex{-2\kappa\cosh x} \\
					&= 2\kappa \ex{2\kappa} K_1(2\kappa) \ ,
\end{align}
where $K_1$ denotes the first modified Bessel function of the second kind.

Since $\kappa \sim (\eps/\Delta\theta)^2$, we expect the strongest-lens approximation to hold only for $\kappa\ll 1$. In that regime, the mean amplification reads
\begin{equation}\label{eq:mean_amplification_s}
\ev{A}\e{s} = 1 + 2\kappa + \mathcal{O}(\kappa^2) \ .
\end{equation}
This result deserves to be noticed, because it relates to the long-standing question of flux conservation in gravitational lensing. As shown by Weinberg in 1976~\cite{1976ApJ...208L...1W}, the mean amplification caused by a very sparse set of isolated, Poisson-distributed, point lenses is identical to the amplification which would be observed if the matter constituting these lenses was homogeneously distributed in space. Equation~\eqref{eq:mean_amplification_s} generalizes Weinberg's result to a nonhomogeneous distribution of lenses.

To understand that claim, it is useful to rewrite $\kappa$ by substituting the expression~\eqref{eq:Einstein_radius_physical} of $r\e{E}$ into Eq.~\eqref{eq:optical_depth}; we obtain
\begin{align}
\kappa &= 4\pi G \int_0^{z\e{s}} \dd z \;
						\frac{D\e{d}(z)D\e{ds}(z)}{D\e{s}} 
						\int \dd m \; m \,
						\frac{\dd\Sigma}{\dd z \dd m} \\
		&= 4\pi G \int_0^{z\e{s}} \dd z \;
						\frac{D\e{d}(z)D\e{ds}(z)}{D\e{s}} \,
						\frac{\rho\e{l}(z)}{(1+z)H(z)} \ ,
\end{align}
where $\rho\e{l}(z)$ is the contribution of the point lenses to the mean mass density\footnote{We are not necessarily talking about a homogeneous Universe here; in particular, $\rho\e{l}(z)\not\propto(1+z)^{-3}$ in general.} at $z$, while a factor $(1+z)H(z)=\dd z/\dd d$ appears to convert redshifts into proper distances. Furthermore, assuming that angular-diameter distances $D\e{d}, D\e{ds}, D\e{s}$ can be computed as in a homogeneous and isotropic Friedmann-Lema\^itre-Robertson-Walker (FLRW) Universe, we find
\begin{equation}\label{eq:interpretation_tau}
\kappa =  \frac{3}{2}\, H_0^2
			\int_0^{\chi\e{s}} \dd \chi \;
		\frac{f_K(\chi)f_K(\chi\e{s}-\chi)}{f_K(\chi\e{s})} \,
		\frac{\rho\e{l}(z)}{(1+z)^2}
\end{equation}
where $\chi$ denotes comoving distances, $K$ is the Universe's spatial curvature parameter, and $f_{K}(\chi)\define\sin(\sqrt{K}\chi)/\sqrt{K}$. Equation~\eqref{eq:interpretation_tau} is the expression of the weak-lensing convergence which would be due to our point lenses, \emph{if their mass were smoothly distributed}, instead of being concentrated into compact objects. In that case, we would have
\begin{equation}
\ev{A}\e{smooth} = 1 + 2\kappa + \mathcal{O}(\kappa^2) \ ,
\end{equation}
which, indeed, agrees with Eq.~\eqref{eq:mean_amplification_s} in the limit $\kappa\ll 1$.

In Ref.~\cite{1986MNRAS.223..113P}, Peacock proposed to heuristically extend the applicability of $p\e{s}(A)$ to higher optical depths, $\kappa\sim 1$, by conjecturing flux conservation. The idea consists in treating the optical depth involved in $p\e{s}(A)$ as a free parameter, denoted $\kappa\e{s}$, and fix it by ensuring that $\ev{A}\e{s}(\kappa\e{s})$ produces the true mean amplification, i.e.\footnote{Peacock seems to have assumed $\ev{A}\e{true}=(1-\kappa)^{-2}$, because he found, e.g., $\kappa\e{s}=199$ for $\kappa=0.8$. This choice delivered results in good agreement with the numerical simulations of Paczy\'{n}ski~\cite{1986ApJ...301..503P}. However, the results of the present article do not support this conclusion; in Sec.~\ref{sec:numerics} we rather find $\ev{A}\e{sim}\approx (1+\sqrt{1+4\kappa})^2/4$, although the discrepancy shall be explained by the finite extension of our simulated map.}
\begin{equation}\label{eq:Peacock}
\ev{A}\e{s}(\kappa\e{s}) = 2\kappa\e{s} \ex{2\kappa\e{s}} K_1(2\kappa\e{s}) = \ev{A}\e{true} \ .
\end{equation}
As will be seen in Sec.~\ref{sec:numerics}, this procedure would not succeed to capture the features of $p(A)$ due to collective effects for $\kappa\sim 1$.

%%%%%%%%%%%%%%%%%%%%
\section{Multiplicative amplifications}
\label{sec:multiplicative}
%%%%%%%%%%%%%%%%%%%%

The strongest-lens approximation is arguably crude, because it completely neglects the long-distance effects due to the other lenses. In order to allow for all the lenses together, one may conjecture that amplifications are multiplicative~\cite{1983ApJ...267..488V}; namely, the total amplification caused by two lenses reads~$A=A_1 A_2$, where $A_1, A_2$ are the individual amplifications that each lens would produce in the absence of the other.

\subsection{Can we really multiply amplifications?}

An intuitive justification of the multiplicative model is the following. Suppose that the two lenses are well separated along the line of sight, lens 1 being the closest to the observer and lens 2 the closest to the source. Let $\vect{\theta}$ be an image. Each lens is individually endowed with an image-to-source mapping of the form~\eqref{eq:lens_eq_1PL}, $\vect{\beta}_1(\vect{\theta}), \vect{\beta}_2(\vect{\theta})$. Considering their successive effect, and assuming that, for lens 2, $\vect{\beta}_1(\vect{\theta})$ plays the role of an intermediate image, the combined mapping would be $\vect{\beta}(\vect{\theta})=\vect{\beta}_2(\vect{\beta}_1(\vect{\theta}))$. The chain rule then yields the amplification as
\begin{equation}
A^{-1}
= \det\pd{\vect{\beta}}{\vect{\theta}}
= \det\pd{\vect{\beta}_2}{\vect{\theta}}\pd{\vect{\beta}_1}{\vect{\theta}}
= (A_2 A_1)^{-1} \ .
\end{equation}

Leaving aside the important fact that $A_2$ must be evaluated at a non-trivial intermediate image position, \emph{the above reasoning is incorrect anyway}. The reason is that gravitational lensing is a somewhat non-local phenomenon: the properties of a given lens depend on everything that happens to light before and after it. For example, the exact lens map corresponding to the combination of two point lenses is
\begin{equation}\label{eq:lens_eq_2_lenses}
\vect{\beta}(\vect{\theta})
= \vect{\theta}
	- \frac{\eps_1^2}{\vect{\theta}-\vect{\lambda}_1}
	- \frac{\eps_2^2}{\vect{\theta}-\vect{\lambda}_2-\frac{\eps_{12}^2}{\vect{\theta}-\vect{\lambda}_1}} \ ,
\end{equation}
where $\eps_1, \eps_2$ are the Einstein radii of the two lenses, $\vect{\lambda}_1, \vect{\lambda}_2$ their positions, and $\eps_{12}\define \sqrt{4G m_1 D_{\text{d}_1\text{d}_2}/D_{\text{o}\text{d}_1} D_{\text{o}\text{d}_2}}$. Equation~\eqref{eq:lens_eq_2_lenses} must be compared with
\begin{equation}
\vect{\beta}_2(\vect{\beta}_1(\vect{\theta}))
= \vect{\theta}
	- \frac{\eps_1^2}{\vect{\theta}-\vect{\lambda}_1}
	- \frac{\eps_2^2}{\vect{\theta}-\vect{\lambda}_2-\frac{\eps_{1}^2}{\vect{\theta}-\vect{\lambda}_1}} \ .
\end{equation}
Since $\eps_1\not= \eps_{12}$, the ansatz $\vect{\beta}_2(\vect{\beta}_1(\vect{\theta}))$ does not properly account for \emph{lens-lens coupling}. This fundamentally prevents amplifications from being multiplicative. Another, equally important, obstacle, is that one must sum the individual amplifications of each image of a given source, which requires to determine these images in the first place.

Of course, when the multiplicative model was first proposed by Ref.~\cite{1983ApJ...267..488V} in 1983, its authors were aware that it was only an approximation. A year later, a major aspect of lens-lens coupling was emphasized by Refs.~\cite{1984A&A...132..168C, 1984JApA....5..235N}, namely shear. If a light source is (even weakly) sheared by a first lens, then the effect of a second lens turns out to be quite different from the no-shear case. In particular, the geometry of its caustics is significantly affected, as well as the amplification statistics. This problem has been thoroughly investigated in the 1980s-1990s, both analytically and numerically~\cite{1986A&A...166...36K, 1987ApJ...319....9S, 1992ApJ...389...63M, 1997ApJ...489..508K, 1997ApJ...489..522L}. More recently, this coupling between the weak-lensing effect of the cosmic web with the properties of strong lenses~\cite{Schneider:1997bq, Birrer:2016xku} has been proposed as a possible measure of cosmic shear using Einstein rings~\cite{Birrer:2017sge}.

\subsection{Practical interest of multiplicativity}
\label{subsec:practical_mutiplicativity}

Albeit inexact, the multiplicative model is practically very convenient, as far as statistics are concerned. Let two lenses, or groups of lenses, generate random amplifications $A_1, A_2$. Assuming multiplicativity, the total amplification reads $\ln A = \ln A_1 + \ln A_2$. Hence, if $L\define \ln A$, then the PDF of $L$ reads
\begin{equation}\label{eq:PDF_L}
P(L) = \int_0^L \dd L_1 \; P_{12}(L_1, L-L_1) \ ,
\end{equation}
where $P_{12}(L_1, L_2)$ is the joint PDF of the individual logarithmic amplifications. If, furthermore, $A_1$ and $A_2$ are independent, then $P_{12}(L_1, L_2)=P_1(L_1) P_2(L_2)$, so that Eq.~\eqref{eq:PDF_L} becomes a convolution product, $P=P_1 * P_2$. This equality can be translated in terms of amplifications, using that
\begin{equation}
p(A) = \abs{\ddf{L}{A}} P(L) = A^{-1} P(\ln A)  \ ,
\end{equation}
which yields
\begin{equation}\label{eq:combining_p_A}
p(A) = \int_1^A \frac{\dd A_1}{A_1} \; p_1(A_1) \, p_2(A/A_1) \ .
\end{equation}
Equation~\eqref{eq:combining_p_A} is a simple prescription to combine independent multiplicative amplifications, which will be useful for the remainder of this section.

\subsection{Derivation of the amplification PDF}
\label{subsec:derivation_mutiplicative}

Assuming multiplicativity, an exact expression for the amplification PDF, $p\e{m}(A)$ (subscript ``m'' stands for ``multiplicative'') was derived in 1993 by Pei~\cite{1993ApJ...403....7P}, hereafter \Pei. However, this derivation contains a few errors and inaccuracies---we refer the curious reader to Appendix~\ref{app:Pei} where Pei's method is reproduced and commented. This encouraged us to propose an alternative proof in the present section. Fortunately, the final result is left unchanged.

\subsubsection{From 1 to $N$ lenses in a finite Universe}
\label{subsubsec:derivation_p_m_A}

Consider the same setup as described in Sec.~\ref{subsubsec:setup}, with $N$ lenses in a finite tubular Universe. As suggested by the discussion of Sec.~\ref{subsec:practical_mutiplicativity}, when amplifications are multiplicative, it is more convenient to work with their logarithm. Let $P_N(L)$ be the PDF of $L\define \ln A = \ln A_1 + \ldots \ln A_N$, we then have
\begin{equation}\label{eq:P_N_convolution}
P_N(L) = (P_1 * P_1 * \ldots * P_1)(L) \define P_1^{*N}(L) \ ,
\end{equation}
where $P_1(\ln A)=A p_1(A)$ logarithmic amplification PDF for a single lens. This quantity can be directly deduced from the expression~\eqref{eq:p_1_u_result} of $p_1(u)$, namely,
\begin{equation}
p_1(A)
= \abs{\ddf{u}{A}} p_1(u)
= \frac{2}{(A^2-1)^{3/2}} \, \frac{\ev[1]{r\e{E}^2}}{R^2} \ ,
\end{equation}
as long as $A\geq A\e{c}\define A(u\e{c})$, where $u\e{c}$ is given by Eq.~\eqref{eq:u_c}. Again, it is not necessary to explicitly determine the case $A\leq A\e{c}$ in order to proceed.

The convolution product~\eqref{eq:P_N_convolution} is more easily handled in Fourier space, where it becomes a regular product. We adopt the following convention for Fourier transforms:
\begin{align}
\tilde{P}(K) &= \int_{-\infty}^{\infty} \dd L \; \ex{-\i K L} P(L) \ , \\
P(L) &= \int_{-\infty}^{\infty} \frac{\dd K}{2\pi} \; \ex{\i K L} \tilde{P}(K) \ ,
\end{align}
so that $\tilde{P}_N(K)=\tilde{P}_1^N(K)$. Therefore, by taking the inverse Fourier transform, and restoring the normal amplification variable~$A=\exp L$, we obtain the formal expression 
\begin{equation}\label{eq:p_N}
p_N(A) = \frac{1}{A} \int_{-\infty}^{\infty} \frac{\dd K}{2\pi} \; A^{\i K} 
									\pac{ \int_1^\infty \dd A' \; (A')^{-\i K} p_1(A')}^N .
\end{equation}

\subsubsection{An infinity of lenses}
\label{subsubsec:p_m_infinity}

Let us finally determine the limit $p\e{m}(A)$ of $p_N(A)$ as both the number $N$ of lenses and the size $R$ of the tube go to infinity. For that purpose, the first step consists in manipulating the logarithmic Fourier transform of $p_1$ as
\begin{align}
\label{eq:P_tilde_1_start}
\tilde{P}_1(K)
&= \int_1^\infty \dd A \; A^{-\i K} p_1(A) \\
\label{eq:P_tilde_1_split}
&= \int_1^{A\e{c}} \dd A \; A^{-\i K} p_1(A) + \int_{A\e{c}}^{\infty} \dd A \; A^{-\i K} p_1(A)
\end{align}
For $R\rightarrow\infty$, we have $u\e{c}\rightarrow\infty$, and hence $A\e{c}\rightarrow 1$. Specifically,
\begin{equation}
A\e{c}-1 \sim \frac{2}{u\e{c}^4} = \frac{8 G m\e{max}\mathcal{D}\e{max}}{R^4} \ .
\end{equation}
Hence, the first integral of Eq.~\eqref{eq:P_tilde_1_split} can be expanded as
\begin{align}
\int_1^{A\e{c}} \dd A \; A^{-\i K} p_1(A)
&= \int_1^{A\e{c}} \dd A \; p_1(A) + \mathcal{O}(A\e{c}-1) \\
&= 1 - \int_{A\e{c}}^\infty \dd A \; p_1(A) + \mathcal{O}(R^{-4}) \ ,
\end{align}
where, in the second line, we used the normalization of $p_1(A)$. Substituting the expression of $p_1(A\geq A\e{c})$, we find
\begin{equation}\label{eq:P_tilde_1_almost}
\tilde{P}_1(K)
= 1 + \frac{2\ev[1]{r\e{E}^2}}{R^2} \int_{A\e{c}}^\infty \dd A \; \frac{A^{-\i K}-1}{(A^2-1)^{3/2}} + \mathcal{O}(R^{-4}) \ .
\end{equation}

The lower limit $A\e{c}$ of the integral of Eq.~\eqref{eq:P_tilde_1_almost} can actually be replaced with $1$, because
\begin{equation}
\int_1^{A\e{c}} \dd A \; \frac{A^{-\i K}-1}{(A^2-1)^{3/2}}
\sim -\frac{\i K}{\sqrt{2}} \int_1^{A\e{c}} \frac{\dd A}{\sqrt{A-1}}
= \mathcal{O}(R^{-2}) \ .
\end{equation}
This leaves us with the analytically solvable integral
\begin{equation}
\int_1^\infty \dd A \; \frac{A^{-\i K}-1}{(A^2-1)^{3/2}}
= 1 - \sqrt{\pi}\, \frac{\Gamma(1+\i K/2)}{\Gamma(1/2+\i K/2)} \ ,
\end{equation}
which does not depend on $N,R$.

The last step consists in raising $\tilde{P}_1(K)$ to the $N$th power, which yields
\begin{multline}
\tilde{P}_1^N(K) = \exp\paac{ \frac{2 N\ev[1]{r\e{E}^2}}{R^2} 
													\pac{ 1 - \sqrt{\pi} \frac{\Gamma(1+\i K/2)}{\Gamma(1/2+\i K/2)} } 
												} \\
+ \mathcal{O}(N R^{-4}) \ .
\end{multline}
We recognize the optical depth $\kappa = N\ev[1]{r\e{E}^2}/R^2$, while the remainder $\mathcal{O}(N/R^4)$ goes to zero as $N,R\rightarrow\infty$. Substituting that result into Eq.~\eqref{eq:p_N}, we conclude that
\begin{empheq}[box=\fbox]{equation}
p\e{m}(A)
%&= \lim_{N\rightarrow\infty} p_N(A) \\
= \ex{2\kappa}
\int_{-\infty}^{\infty} \frac{\dd K}{2\pi} \; A^{\i K-1}
\exp\pac{ -2\kappa \sqrt{\pi}\, \frac{\Gamma(1+\i K/2)}{\Gamma(1/2+\i K/2)} } ,
\label{eq:p_A_result}
\end{empheq}
where $\Gamma$ denotes the usual Gamma function. The above Eq.~\eqref{eq:p_A_result} is equivalent to Eq.~(29) in \Pei, although the present definition~\eqref{eq:optical_depth} of $\kappa$ is slightly more general, because it allows the lenses to have different masses, and to be inhomogeneously distributed along the line of sight.

\subsection{Discussion: low-optical-depth behavior and moments}

Albeit complicated due to its highly oscillatory integrand, the expression~\eqref{eq:p_A_result} of $p\e{m}(A)$ has a simple limit in the low-optical depth limit~$\kappa\ll 1$. Indeed, expanding the exponential, we have
\begin{align}
p\e{m}(A)
&= \frac{2\kappa}{A} \int_{-\infty}^\infty \frac{\dd K}{2\pi} \; \; A^{\i K}
									\pac{ 1 - \sqrt{\pi}\, \frac{\Gamma(1+\i K/2)}{\Gamma(1/2+\i K/2)} }
	+ \mathcal{O}(\kappa^2) \\
\label{eq:asymptotics_p_m_calculation}
&= \frac{2\kappa}{A} \int_{-\infty}^\infty \frac{\dd K}{2\pi} \;
									\int_1^\infty \dd A' \; \frac{(A/A')^{\i K}-A^{\i K}}{[(A')^2-1]^{3/2}}
	+ \mathcal{O}(\kappa^2) \\
&= \frac{2\kappa}{(A^2-1)^{3/2}} + \mathcal{O}(\kappa^2) \qquad \text{if $A\not= 0$},
\end{align}
in agreement with \Pei. The latter result is obtained by using that, in Eq.~\eqref{eq:asymptotics_p_m_calculation}, integration over $K$ yields an integrand proportional to $\delta(\ln A-\ln A') - \delta(\ln A)$. Note that, as expected, this low-$\kappa$ behavior exactly coincides with the one of $p\e{s}(A)$. It is not trivial, however, to which extent the $\mathcal{O}(\kappa^2)$ terms can be neglected, especially for $A-1\ll 1$, because $(A^2-1)^{3/2}$ is nonintegrable on $(1,\infty)$.

The moments of $p\e{m}(A)$ are more conveniently determined starting from Eq.~\eqref{eq:P_tilde_1_start}, replacing $A^{-\i K}$ with $A^n$, and then following the same calculation as in Sec.~\ref{subsubsec:p_m_infinity}. The result is
\begin{equation}
\ev{A^n}\e{m} = \exp \pac{ 2\kappa \int_1^\infty \dd A \; \frac{A^n-1}{(A^2-1)^{3/2}} }\ .
\end{equation}
For $n\geq 2$, the integral diverges because of the upper limit $A\rightarrow\infty$. The mean amplification is, therefore, the only nonzero moment of $p\e{m}(A)$, and reads
\begin{equation}\label{eq:mean_amplification}
\ev{A}\e{m} = \ex{2\kappa} \ ,
\end{equation}
again in agreement with \Pei.

For $\kappa\ll 1$, we find $\ev{A}\e{m} = 1+2\kappa+\mathcal{O}(\kappa^2)$ just like in the strongest-lens model. It is instructive to further compare the behaviors of $\ev{A}\e{s}$ and $\ev{A}\e{m}$; in particular,
\begin{equation}
\ev{A}\e{s} \leq 1+2\kappa \leq \ev{A}\e{m} \ ,
\end{equation}
anticipating on the results of Fig.~\ref{fig:Amean}. These inequalities are a hint that the strongest-lens model may systematically underestimate amplifications, while the multiplicative model may systematically overestimate them.

\subsection{Improving the multiplicative model: a hybrid approach}

One reason why the multiplicative model overestimates amplifications may be that it virtually accounts for two images per lens, which is not realistic. In order to simply allow for the combined effect of many lenses, without dramatically overestimating it, we propose the following hybrid approach between the strongest-lens and multiplicative model.

\subsubsection{Hybrid model}

Suppose that a light source is strongly affected by one lens, and weakly affected by all the others. Let us assume that each weak lens produces a single image---the principal image, with amplification~$A_+$. Let us also neglect lens-lens coupling, so that the amplification~$A\e{s}$ of the strong lens is left unchanged. Then, the total amplification reads~$A\e{s} A\e{w}^+$, where $A\e{w}^+$ is the product of all the principal amplifications of the weak lenses.

Furthermore, let us adopt a mean-field approach\footnote{It is actually possible to proceed without this simplifying assumption. However, the final analytic result is too complicated to have any practical interest. We thus restrict to the mean-field case in this section.}, and replace the stochastic contribution~$A\e{w}^+$ by its statistical average $\bar{A}\e{w}^+$. Our hybrid model is then defined by
\begin{equation}
A = A\e{s} \, \bar{A}\e{w}^+(A\e{s}) \ ,
\end{equation}
where $\bar{A}\e{w}^+$ depends on $A\e{s}$ because of the constraint that the lenses contributing to $A\e{w}$ are weaker than the main strong lens producing~$A\e{s}$. In this model, the stochasticity of $A$ is entirely controlled by the strongest amplification~$A\e{s}$. In other words, the amplification PDF of the hybrid model reads
\begin{equation}
p\e{h}(A) = \ddf{A\e{s}}{A} \, p\e{s}(A\e{s}) \ .
\end{equation}
The remaining task thus consists in computing $\bar{A}\e{w}^+(A\e{s})$.

\subsubsection{Calculation of the mean weak amplification}

Let us consider again the finite setup described in Sec.~\ref{subsubsec:setup}. Suppose that the strongest lens has a reduced impact parameter~$u\e{s}$. Then all the other lenses must satisfy $u\geq u\e{s}$, so that for any one of them
\begin{equation}
p_1(m,z,u|u\e{s}) = \frac{2 u\,r\e{E}^2}{R^2-u\e{s}^2 r\e{E}^2} \, \Theta(R-u r\e{E}) \, \Theta(u-u\e{s}) \ .
\end{equation}
which can be marginalized over $m,z$ to get $p_1(u|u\e{s})$. The result can then be converted in terms of amplifications using Eq.~\eqref{eq:A_pm_u}, and we find
\begin{multline}
p_1(A_+|A\e{s}) = \frac{1}{[A_+(A_+-1)]^{3/2}} \ev{\frac{r\e{E}^2}{R^2-u\e{s}^2 r\e{E}^2}} \\
					\times \Theta(A_+-A\e{c}^+) \, \Theta\pa{ \frac{1+A\e{s}}{2} - A_+} \ ,
\end{multline}
where $A\e{c}^+ = A_+(u\e{c})$, $u\e{c}$ being given by Eq.~\eqref{eq:u_c} to ensure that $b\leq R$. The second Heaviside function corresponds to $u\leq u\e{s}$; note that this is not equivalent to $A_+\leq A\e{s}$ because $A_+(u) = 1/2+A(u)/2 \not= A(u)$. The mean weak amplification due to a single lens besides the strongest lens thus reads
\begin{equation}
\ev{A_+} = \frac{\ev[1]{r\e{E}^2}}{R^2} \int_{A\e{c}}^{(1+A\e{s})/2} \frac{A_+ \, \dd A_+}{[A_+(A_+-1)]^{3/2}}
					+ \mathcal{O}(R^{-4}) \ .
\end{equation}

In the finite tubular Universe, the average weak amplification reads~$\bar{A}\e{w}^+ = \ev{A_+}^N$. In the limit $N, R\rightarrow\infty$, following a similar computation as in Sec.~\ref{subsubsec:p_m_infinity}, we finally get
\begin{align}
\bar{A}\e{w}^+(A\e{s})
&= \exp\pa{ \kappa \int_1^{(1+A\e{s})/2} \dd A_+ \; \frac{A_+-1}{[A_+(A_+-1)]^{3/2}} } \\
&= \exp\pa{ \kappa \sqrt{\frac{A\e{s}-1}{A\e{s}+1}} } \ .
\end{align}
Summarizing, in the hybrid model the amplification PDF reads
\begin{empheq}[box=\fbox]{align}
p\e{h}(A)
&= \ddf{}{A} \, \exp\pac{ -2\kappa \pa{ \frac{A\e{s}}{\sqrt{A\e{s}^2-1}} - 1} }\\
&= \frac{2\kappa \exp\paac{ -\kappa \pac{ (3A\e{s}-1) / \sqrt{A\e{s}^2-1} - 2}} }
										{(A\e{s}^2-1)^{3/2} + \kappa A\e{s}(A\e{s}-1)} \ ,\\
\text{with} \quad
A &= A\e{s} \exp\pa{ \kappa \sqrt{\frac{A\e{s}-1}{A\e{s}+1}} } \ .
\end{empheq}
In practice, one has to numerically invert the relation~$A(A\e{s})$ in order to compute $p\e{h}(A)$.

%\subsubsection{Without mean field}
%
%If $A=A\e{s}A\e{w}$, then
%%
%\begin{equation}
%p\e{h}(A) = \int_0^A \dd A\e{s} \; p\e{s}(A\e{s}) \, p\e{w}(A/A\e{s}|A\e{s}) \ ,
%\end{equation}
%%
%where $p\e{w}(A\e{w}|A\e{s})$ is the PDF of the amplification given by all the weak lenses. It reads
%%
%\begin{equation}
%p\e{w}(A\e{w}|A\e{s})
%= \frac{1}{A\e{w}} \int_{-\infty}^\infty \frac{\dd K}{2\pi} \; A\e{w}^{\i K} \, \ex{\kappa I_+(K, A\e{s})}
%\end{equation}
%%
%where $I_+(K,A\e{s})$ is an integral which can be expressed in terms of the hypergeometric function as
%%
%\begin{align}
%I_+
%&= \int_1^{(1+A_0)/2}
%		\frac{\dd A_+}{2} \; \frac{A_+^{-\i K}-1}{[A_+(A_+-1)]^{3/2}} \\
%&= \frac{2 A\e{s}}{\sqrt{A\e{s}^2-1}}
%	-\sqrt{\frac{2}{A\e{s}-1}}  \,
%	\tensor[_2]{F}{_1}\pa{ -\frac{1}{2}, \frac{3}{2}+\i K; \frac{1}{2}; \frac{1-A\e{s}}{2}} .
%\end{align}
%%
%The resulting PDF is prohibitively complicated to calculate numerically (one would better perform a ray-tracing calculation\ldots)

\subsubsection{Asymptotic behavior and moments}

In the low-optical-depth regime, $A\e{s}=A+\mathcal{O}(\kappa)$, and hence we find again
\begin{equation}
p\e{h}(A) = \frac{2\kappa}{(A^2-1)^{3/2}} + \mathcal{O}(\kappa^2) \ ,
\end{equation}
just like $p\e{s}, p\e{m}$. The amplitude of the high-amplification tail is also easily obtained whatever $\kappa$: for $A\gg 1$, $A\approx \ex{\kappa}A\e{s}$, so that $p\e{h}(A\gg 1) \approx 2\kappa \ex{2\kappa}/A^3 = \ex{2\kappa}p\e{s}(A\gg 1)$. Thus, in the optically-thick regime, the tail is significantly higher than for the strongest-lens approximation.

Just like in the other approximation schemes, the $A^{-3}$ algebraic tail for large amplifications implies that only the first moment of $p\e{h}(A)$, i.e. the mean amplification, is finite. It can be computed analytically as
\begin{align}
\ev{A}\e{h}
&\define \int_1^\infty \dd A \; A \, p\e{h}(A)  \\
\label{eq:A_h_calculation_1}
&= \int_1^\infty \dd A\e{s} \; A\e{s} \, \exp\pa{\kappa\sqrt{\frac{A\e{s}-1}{A\e{s}+1}}} p\e{s}(A\e{s}) \\
\label{eq:A_h_calculation_2}
&= \kappa \, \ex{2\kappa} \int_1^\infty \dd x \pa{1+\frac{1}{x^2}} \ex{-\kappa x}\\ 
&= \ex{\kappa} + \kappa^2 \ex{2\kappa} \Gamma(-1,\kappa) \ ,
\end{align}
where, from Eq.~\eqref{eq:A_h_calculation_1} to Eq.~\eqref{eq:A_h_calculation_2}, we defined the variable $x=\sqrt{(A\e{s}+1)/(A\e{s}-1)}$, and $\Gamma(a,y)$ denotes an incomplete Gamma function,
\begin{equation}
\Gamma(a,y) \define \int_y^\infty \dd t \; t^{a-1} \, \ex{-t} \ .
\end{equation}

%%%%%%%%%%%%%%%%
\section{Extended sources}
\label{sec:extended_sources}
%%%%%%%%%%%%%%%%

Hitherto, all the calculations were performed assuming that the light sources undergoing microlensing were point-like. However, it is quite instructive to investigate the case where this assumption is relaxed, especially if one is interested about the collective effect of many low-mass lenses. Seminal works on that specific topic include Refs.~\cite{1986A&A...157..383N, 1987A&A...171...49S, 1987A&A...179...71S, 1991A&A...251..393M}.

In full generality, an extended source~$\mathcal{S}$ can be modelled as a collection of infinitesimal patches. The apparent luminous intensity\footnote{We call luminous intensity the electromagnetic power per unit area received by the detector.} of such an infinitesimal patch~$\dd^2\vect{\beta}$ of intrinsic intensity $\dd^2 I\e{s}$ being $A(\vect{\beta})\dd^2 I\e{s}$, the total apparent intensity of the extended source reads
\begin{equation}
I = \int_{\mathcal{S}} \dd^2\vect{\beta} \;
	\frac{\dd^2 I\e{s}}{\dd^2\vect{\beta}} \, A(\vect{\beta})\ .
\end{equation}
The quantity $\dd^2 I\e{s}/\dd^2\vect{\beta}$ is called the specific intensity of the source. From the above, we immediately conclude that the net amplification of the extended source reads
\begin{equation}\label{eq:A_ext_general}
A\e{ext}
= \frac{\int_{\mathcal{S}} \dd^2\vect{\beta} \;
			 \frac{\dd^2 I\e{s}}{\dd^2\vect{\beta}} \, A(\vect{\beta})
			}
			{\int_{\mathcal{S}} \dd^2\vect{\beta} \;
				\frac{\dd^2 I\e{s}}{\dd^2\vect{\beta}}
			} \ .
\end{equation}
In other words, accounting for the finite extension of sources is equivalent to \emph{smoothing} amplification maps.

\subsection{Amplification of a disk source by a single one point lens}

\begin{figure}[h!]
\centering
\includegraphics[width=\columnwidth]{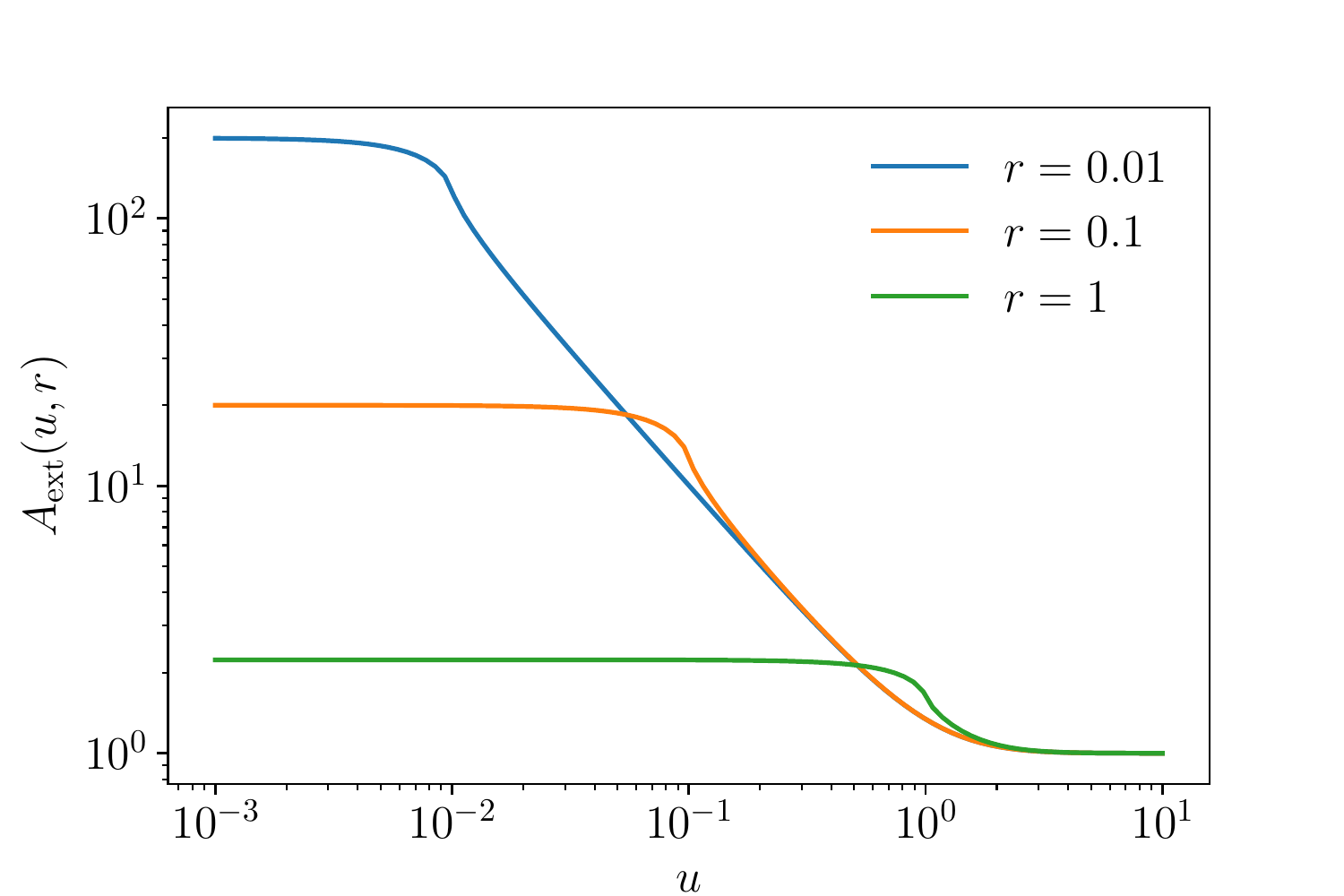}
\caption{Amplification $A\e{ext}(u,r)$ of a disk source by a single point lens with Einstein radius $\eps$, as a function of the reduced impact parameter $u=\beta/\eps$ and the reduced radius of the source $r=\sigma/\eps$.}
\label{fig:Aext}
\end{figure}

Let us assume, for simplicity, that $\mathcal{S}$ is a disk with apparent angular radius $\sigma$, and whose specific intensity is homogeneous. In that case, Eq.~\eqref{eq:A_ext_general} simplifies into
\begin{equation}
A\e{ext} \define \frac{1}{\pi \sigma^2}\int_{\mathcal{S}} \dd^2\vect{\beta} \; A(\vect{\beta}) \ .
\end{equation}
If the source is lensed by a single point lens, an analytical expression for $A\e{ext}$ was found by Ref.~\cite{1994ApJ...430..505W} as below, and represented in Fig.~\ref{fig:Aext}
\begin{multline}\label{eq:A_ext_disk}
A\e{ext}(u,r)
= \frac{u+r}{2\pi r^2} \sqrt{4+(u-r)^2} \, \mathrm{E}(m) \\
	- \frac{u-r}{2\pi r^2} \frac{8+(u^2-r^2)}{\sqrt{4+(u-r^2)}} \, \mathrm{F}(m) \\
	+ \frac{2(u-r)^2}{\pi r^2(u+r)} \frac{1+r^2}{\sqrt{4+(u-r)^2}} \, \Pi(n,m) \ ,
\end{multline}
where, as before, $u$ denotes the reduced impact parameter of the source; the new parameter $r\define \sigma/\eps$ describes the size of the source in units of the lens' Einstein radius. The functions $\mathrm{E}, \mathrm{F}, \Pi$ are the complete elliptic integrals of the first, second, third type, respectively, and
\begin{equation}
n\define \frac{4ur}{(u+r)^2} \ ,
\qquad
m \define \frac{4n}{4+(u-r)^2} \ .
\end{equation}
Note that, in the above expressions, we used Wolfram's convention\footnote{\href{https://www.wolframalpha.com/input/?i=ellipticE}{\tt https://www.wolframalpha.com/input/?i=ellipticE}} for elliptic integrals, which differs from the Gradshteyn \& Ryzhik convention used in Ref.~\cite{1994ApJ...430..505W}.

The point-source case is shown to be recovered for $r\rightarrow 0$. Contrary to what happens in that case, when $r\not=0$, the amplification is bounded up. The maximal amplification happens when $u=0$, and reads
\begin{equation}
A\e{max}(r) \define A\e{ext}(0,r) = \sqrt{1+\frac{4}{r^2}} \ .
\end{equation}
As expected, $A\e{max}$ diverges for $r\rightarrow 0$.

\subsection{Impact on amplification statistics}

Let us now investigate how the finite extension of sources affect amplification statistics. We will restrict that analysis to the strongest-lens and multiplicative models.

\subsubsection{In the strongest-lens model}
\label{subsubsec:p_s_A_ext}

The computation of $p(A)$ with extended sources can be adapted from the point-source derivation of Sec.~\ref{subsec:strongest_lens}. In doing so, the main difficulty is that $A$ is no longer a function of the reduced impact parameter~$u$ only. We must now account for the dependence in $r=\sigma/\eps$.

Let us start with the simplified case where all the lenses have the same Einstein radius~$\eps$. Then $r$ only takes a single value, and hence all the steps of Sec.~\ref{subsec:strongest_lens} can be repeated identically, except that $A(u)$ must be replaced by $A\e{ext}(u,r)$. This yields
\begin{equation}\label{eq:p_s_ext_A_single_eps}
p\e{s}\h{ext}(A,\eps) = \pd{}{A} \ex{-\kappa u^2(A, \sigma/\eps)} \ ,
\end{equation}
where $A\mapsto u(A,r)$ is the inverse of $u\mapsto A\e{ext}(u,r)$. Unfortunately, even in the special case of a homogeneous disk-source, Eq.~\eqref{eq:A_ext_disk} cannot be inverted analytically.

How is that result generalized to lenses with a distribution of Einstein radii? What prevents us from following the derivation of Sec.~\ref{subsec:strongest_lens} is that the strongest lens is no longer the lens with the smallest $u$; indeed, if the latter turns out to be characterized by a large ratio $r$, its amplification may be significantly damped, so that another lens with higher $u$ may be more efficient.

Nevertheless, this difficulty can be circumvented as follows. Let us organize the lenses in groups characterized by a given Einstein radius. Each group $i$ is thus equipped with an $\eps_i$ (hence an $r_i=\sigma/\eps_i$), and its importance is quantified by a partial optical depth $\kappa_i$. The strongest amplification is then the strongest of the strongest within each group, and hence the associated PDF reads
\begin{align}
p\e{s}\h{ext}(A)
&= \sum_i p\e{s}\h{ext}(A,\eps_i) \,
		\prod_{j\not=i} \int_1^A \dd A' \; p\e{s}\h{ext}(A',\eps_j)\\
&= \sum_i \kappa_i \pd{u^2(A,r_i)}{A} \,
		\exp\pac{- \sum_{j=1}^{N\e{p}} \kappa_j u^2(A,r_j)} \\
&= \ddf{}{A} \, \ex{-\kappa u^2\e{eff}(A)} \ ,
\end{align}
where $\kappa$ is the sum of the partial optical depths of all the groups of lenses, and we introduced the optical-depth-weighed impact parameter
\begin{align}\label{eq:u_eff}
u\e{eff}^2(A)
&\define \frac{1}{\kappa} \sum_i \kappa_i u^2(A, \sigma/\eps_i) \\
&= \frac{1}{\kappa} \int \dd \eps \; \ddf{\kappa}{\eps} \, u^2(A, \sigma/\eps)
\end{align}
in the continuous limit.

\subsubsection{In the multiplicative model}

With the multiplicative approach, we have found in Sec.~\ref{subsubsec:derivation_p_m_A} that $p\e{m}(A)$ takes the form of an inverse Fourier transform
\begin{equation}
p\e{m}(A) = \int_{-\infty}^\infty \frac{\dd K}{2\pi} A^{\i K-1} \tilde{P}\e{m}(K) \ ,
\end{equation}
with
\begin{equation}\label{eq:P_tilde_m_one_population}
\tilde{P}\e{m}(K) = \exp\pac{ \kappa \int_1^\infty \dd A \; (A^{-\i K}-1) \abs{\ddf{u^2}{A}} } \ .
\end{equation}
These expressions turn out to be independent of the actual expression of $A(u)$, and thus partly generalize from the point-source case to the extended-source case. Specifically, if all the lenses have the same Einstein radius $\eps$, we simply have
\begin{equation}
\tilde{P}\e{m}\h{ext}(K; \eps, \kappa)
= \exp\pac{ \kappa \int_1^\infty \dd A \; (A^{-\i K}-1) \abs{\pd{u^2(A;r)}{A}} } \ ,
\end{equation}
where, again, $u(A;r)$ denotes the inverse of $A\e{ext}(u;r)$.

For a general population of lenses, we can proceed similarly to Sec.~\ref{subsubsec:p_s_A_ext} and organize them in groups of identical Einstein radii. Due to the multiplicative assumption, the Fourier transform of the logarithmic total amplification reads
\begin{align}
\tilde{P}\e{m}\h{ext}(K)
&= \prod_{i=1}^{N\e{p}} \tilde{P}\e{m}\h{ext}(K;\kappa_i,r_i) \\
&=  \exp\pac[4]{ \kappa \int_1^\infty \dd A \; (A^{-\i K}-1)\, \abs[4]{\ddf{u\e{eff}^2}{A}} } \ ,
\end{align}
where $u\e{eff}^2$ is the same as defined in Eq.~\eqref{eq:u_eff}. Contrary to the point-source case, there is no analytic expression of the above integral. Transforming the integral over $A$ into an integral over $u^2$, we conclude that
\begin{equation}\label{eq:p_m_extended_source}
p\e{m}\h{ext}(A)
= \int_{-\infty}^\infty \frac{\dd K}{2\pi} A^{\i K-1} \exp\paac{ \kappa \int_0^{\infty} \dd u^2 \; [A\e{eff}^{-\i K}(u)-1] } ,
\end{equation}
where $A\e{eff}(u)$ is the inverse of $u\e{eff}(A)$.

\subsection{Discussion}

Just like in the infinitesimal-source case, both the strongest-lens model and the multiplicative model coincide in the limit $\kappa\ll 1$. Expanding the exponential of Eq.~\eqref{eq:p_m_extended_source} at first order in $\kappa$, we find for $A>1$
\begin{equation}\label{eq:p_m=p_s_ES}
p\e{m}\h{ext}(A) =
\kappa \abs{\ddf{u\e{eff}^2}{A}} + \mathcal{O}(\kappa^2)
= p\e{s}\h{ext}(A) \ .
\end{equation}
Beyond that regime, the strongest-lens and multiplicative models have qualitatively distinct behaviors, especially at large amplifications. Indeed, while $p\e{s}\h{ext}(A)$ experiences a sharp drop to zero as $A$ reaches its maximum value, there is no such drop for $p\e{m}\h{ext}(A)$. Instead, the high-$A$ tail of the latter decreases more rapidly than in the infinitesimal-source case, without being totally annihilated.

The above analysis of finite-size effects shows that they are controlled by the dimensionless parameter $r=\sigma/\eps=R\e{s}/r\e{E}$, where $R\e{s}$ is the physical size of the source. It is instructive to evaluate the order of magnitude of $r$ as expected in relevant setups. For a type-Ia supernova at $z=1$ lensed by an object of mass $M$, we find
\begin{equation}
r\e{SN} = 2.0 \times \frac{R\e{s}}{100\U{AU}}
	\sqrt{\frac{10^{-3} M_\odot}{M}}
	\sqrt{\frac{300\U{Mpc}}{\mathcal{D}}} \ ,
\end{equation}
where $\mathcal{D}$ is the distance ratio defined in Eq.~\eqref{eq:Einstein_radius_physical}. In that situation, $r$ becomes comparable to unity for $M\lesssim 10^{-3}M_\odot$. For a quasar at $z=2$, we have
\begin{equation}
r\e{QSO} = 1.6 \times \frac{R\e{s}}{0.5\U{pc}}
	\sqrt{\frac{10^3 M_\odot}{M}}
	\sqrt{\frac{500\U{Mpc}}{\mathcal{D}}} \ ,
\end{equation}
so that finite-size effects already kick in for massive lenses.

%%%%%%%%%%%%%%%%%%%%%%%%%%
\section{Comparison with numerical simulations}
\label{sec:numerics}
%%%%%%%%%%%%%%%%%%%%%%%%%%

How well do the analytic models~$p\e{s}(A)$, $p\e{m}(A)$, or $p\e{h}(A)$ reproduce the actual amplification PDF for a given optical depth $\kappa$? This question can be partially addressed using numerical simulations. Section~\ref{subsec:ray_shooting} introduces the inverse ray-shooting technique used for that purpose; Secs.~\ref{subsec:mean_amplification}, \ref{subsec:comparing_PDF} compare the three models' performance in reproducing respectively the mean amplification and full PDF; and Sec.~\ref{subsec:extended_sources} focuses on the impact of extended sources.

\subsection{Numerical method: inverse ray shooting}
\label{subsec:ray_shooting}

\subsubsection{Principle and setup}

Inverse ray shooting~\cite{1986A&A...166...36K, 2006ApJ...653..942M} is a conceptually simple method to generate amplification maps, PDFs, and lightcurves. Its principle relies on the fact that the lens map expresses the position of the source~$\vect{\beta}$ of a given image $\vect{\theta}$ (whence the name \emph{inverse} ray shooting). Consider $N$ fictitious images~$\vect{\theta}$, arranged on a regular grid\footnote{Another option is to pick image positions randomly, but the resulting shot noise significantly reduces the accuracy of the code.}, with homogeneous surface density $n\e{im}=\dd^2 N/\dd^2\vect{\theta}$. Map these $N$ images  to their respective sources, using $\vect{\beta}(\vect{\theta})$, so that the source plane is inhomogeneously filled with points. Call $n\e{s}(\vect{\beta})=\dd^2 N/\dd^2\vect{\beta}$ its surface density. Since the number of points is conserved during the lens mapping,
\begin{equation}
A(\vect{\beta})
\define \frac{\dd^2 \vect{\theta}}{\dd^2\vect{\beta}}
=\frac{\dd^2 \vect{\theta}}{\dd^2 N} \, \frac{\dd^2 N}{\dd^2\vect{\beta}}
= \frac{n\e{s}(\vect{\beta})}{n\e{im}} \ .
\end{equation}
Thus, the surface density of the fictitious sources directly tells us about the amplification in the source plane.

The simulations presented in this article were performed with identical lenses, randomly distributed on a plane (2D lensing). In that case, the lens map takes the form
\begin{equation}
\vect{\beta}(\vect{\theta}) = \vect{\theta} - \sum_{k=1}^{N\e{l}} \frac{\eps^2}{\vect{\theta}-\vect{\lambda}_k} \ ,
\end{equation}
where $\eps, \vect{\lambda}_k$ are respectively the Einstein radius and angular position of the $k$th lens, and $N\e{l}$ is the total number of lenses. The 2D-lensing choice was mostly made for simplicity, although it would not be difficult to generalize it to a 3D distribution of lenses, e.g., using multi-plane lensing~\cite{1986ApJ...310..568B}. Reference~\cite{1997ApJ...489..522L} found that, for a given optical depth, 3D lensing tends to give rise to \emph{fewer} caustic mergers than 2D lensing. This suggests that the impact of lens-lens coupling is weaker in 3D than in 2D. Therefore, the departure from the predictions of the strongest-lens model that will be described here shall be seen as an upper-bound estimate.

\begin{figure}[h!]
\centering
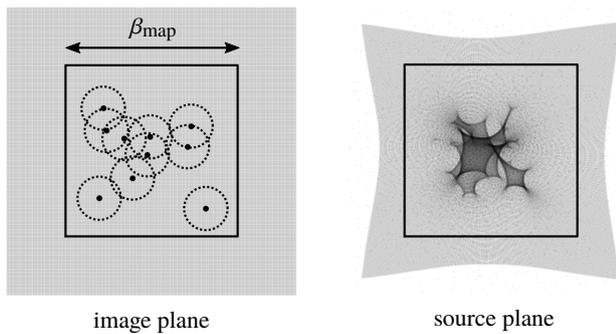
\caption{Example of inverse ray shooting. \textit{Left panel}: Image plane, where $10^5$ fictitious images~$\vect{\theta}$ are regularly arranged on a grid. The position of the lenses are indicated by black dots, and their individual Einstein radii are shown with dashed lines. \textit{Right panel}: Source plane, indicating the positions of the sources~$\vect{\beta}(\vect{\theta})$ of the fictitious images. Darker regions (higher density of points) are regions with higher amplification. The framed square represents the region which we keep for further computations.}
\label{fig:example_ray_shooting}
\end{figure}

In practice, we consider a square map with edge $\beta\e{map}$ in the source plane. Given a number~$N\e{l}$ of lenses, we then choose their Einstein radius~$\eps$ such that the desired optical depth~$\kappa=N\e{l}\pi \eps^2/\beta\e{map}^2$ is reached. The lens positions~$\vect{\lambda}_k$ are then randomly picked in such a way that their individual Einstein disk $\pi\eps^2$ is entirely comprised in the map. An example of inverse ray shooting, with $N\e{l}=10$ and $\kappa=0.5$ is given in Fig.~\ref{fig:example_ray_shooting}. Note the important contraction from the image plane to the source plane, which requires the area of the image plane to be larger than the considered area in the source plane.

\subsubsection{Adaptive mesh refinement}

In order to get a high-resolution amplification map while maintaining good accuracy, it is necessary to generate a very large number of points in the source plane. Then, several options are available to evaluate $n\e{s}(\vect{\beta})$ in the source plane. Kernel density estimation algorithms can be very accurate, but they are computationally too expensive. We chose to simply count the number of sources which end in each pixel of the source plane. However, since amplification (and hence $n\e{s}$) varies over several orders of magnitude throughout the map, it is necessary to adapt the pixel size depending on its position.

\begin{figure}[h!]
\centering
\includegraphics[width=\columnwidth]{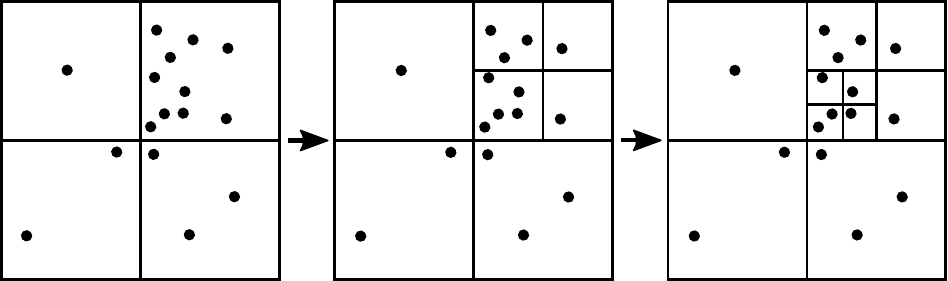}
\caption{Adaptive mesh refinement from a coarse grid to a locally finer grid. In this example, the threshold is $N\e{max}=4$.}
\label{fig:AMR}
\end{figure}

\begin{figure}[h!]
\centering
\includegraphics[width=\columnwidth]{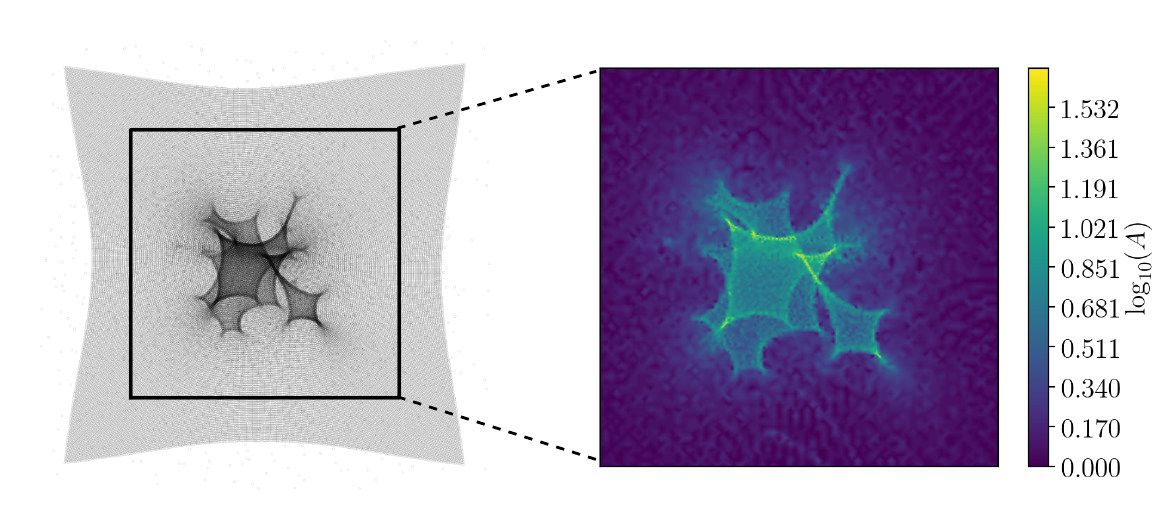}
\caption{Amplification map, in the source plane, generated from the set of sources of the left panel using AMR.}
\label{fig:sources_to_amap}
\end{figure}

We addressed this issue with a simple \emph{adaptive mesh refinement} (AMR) procedure. In cosmology, AMR is particularly useful to resolve high-density regions in $N$-body codes, such as \textsc{Ramses}~\cite{2002A&A...385..337T}. Its principle is depicted in Fig.~\ref{fig:AMR}. Starting from a regular coarse grid, count the number of sources in each pixel; if the number of sources exceeds a given threshold $N\e{max}$, divide the pixel in four equal subpixels, and repeat this operation with the subpixels until all of them contain less than $N\e{max}$ sources. The density of sources in a subpixel of area~$\Omega\e{sub}$ containing $N\e{sub}$ sources is then estimated as $n\e{s}=N\e{sub}/\Omega\e{sub}$. Since $\Omega\e{sub}$ can be very small, AMR allows one to locally access very high values of the amplification without the need to divide the whole map into tiny pixels. An example of amplification map obtained from this procedure is given in Fig.~\ref{fig:sources_to_amap}.

\subsubsection{Performance}

The code has been written in Python. In its current, nonoptimized and nonparallelized form, for $N\e{l}=100$, $N=2\times 10^7$, it takes roughly 6 hours to generate a full map on a laptop with Intel Core i5 CPU. Computing time is linear in $N\e{l} N$. A file containing all source positions typically occupies hundreds of MB to a few GB, depending on $N$. This amount of disk space is divided by 20 once the sources are arranged on a refined map, and their exact positions are deleted.

\subsection{Mean amplification}
\label{subsec:mean_amplification}

As a first application, we consider the mean amplification~$\ev{A}$ over the simulated map. Note that~$\ev{\cdots}$ represents an average over randomly distributed \emph{sources}. This source-averaging procedure must be distinguished from directional averaging, which would give the same weight to random directions in the image plane~\cite{Kibble:2004tm, Kaiser:2015iia, Bonvin:2015kea, Fleury:2016fda}. From a simulated amplification map, such as the one depicted in Fig.~\ref{fig:sources_to_amap}, it is straightforward to estimate $\ev{A}$ as follows:
\begin{equation}
\ev{A}
= \sum_i \frac{\Omega_i}{\beta\e{map}^2} A_i
= \sum_i \frac{\Omega_i}{\beta\e{map}^2} \frac{N_i}{\Omega_i n\e{im}}
= \frac{N\e{s}}{N\e{im}} \ ,
\end{equation}
where the sum runs over each (sub-)pixel~$i$ of the map, $\Omega_i$ being the area of the (sub-)pixel, and $N_i$ the sumber of sources in it. The areas~$\Omega_i$ cancel out, so that $\ev{A}$ is simply the ratio between the total number of sources~$N\e{s}$ in the map, and the number of images which were in the same region before lens mapping.

The average amplification can also be estimated as the relative enhancement of the map's area from the source plane to the image plane, $\ev{A}=\Omega\e{im}/\Omega\e{s}$. Let us approximate the map as a disk with radius~$\beta\e{map}$ instead of a square, and deal with the set of lenses as a single point lens, with squared Einstein radius~$\Theta\e{E}^2=N\e{l}\eps^2=\kappa\beta\e{map}^2$. Then the principal image of the map's contour is a circle with radius $\theta\e{map} = \pa[2]{ \beta\e{map} + \sqrt{\beta\e{map}^2+4\Theta\e{E}^2} }/2$. Thus,
\begin{equation}\label{eq:A_mean_est}
\ev{A}
\approx \pa{\frac{\theta\e{map}}{\beta\e{map}}}^2
= \frac{1}{4} \pa{ 1+\sqrt{1+4\kappa} }^2 \ .
\end{equation}
Note the importance of the \emph{finite extent} of region containing the lenses in the above estimate. This is what allows us to treat the $N\e{l}$ lenses as a single point lens producing an image of the map's edge. If the lens distribution was infinite in extent, the expected result would be $\ev{A}=(1-\kappa)^{-2}$.

\begin{figure}[h!]
\centering
\includegraphics[width=\columnwidth]{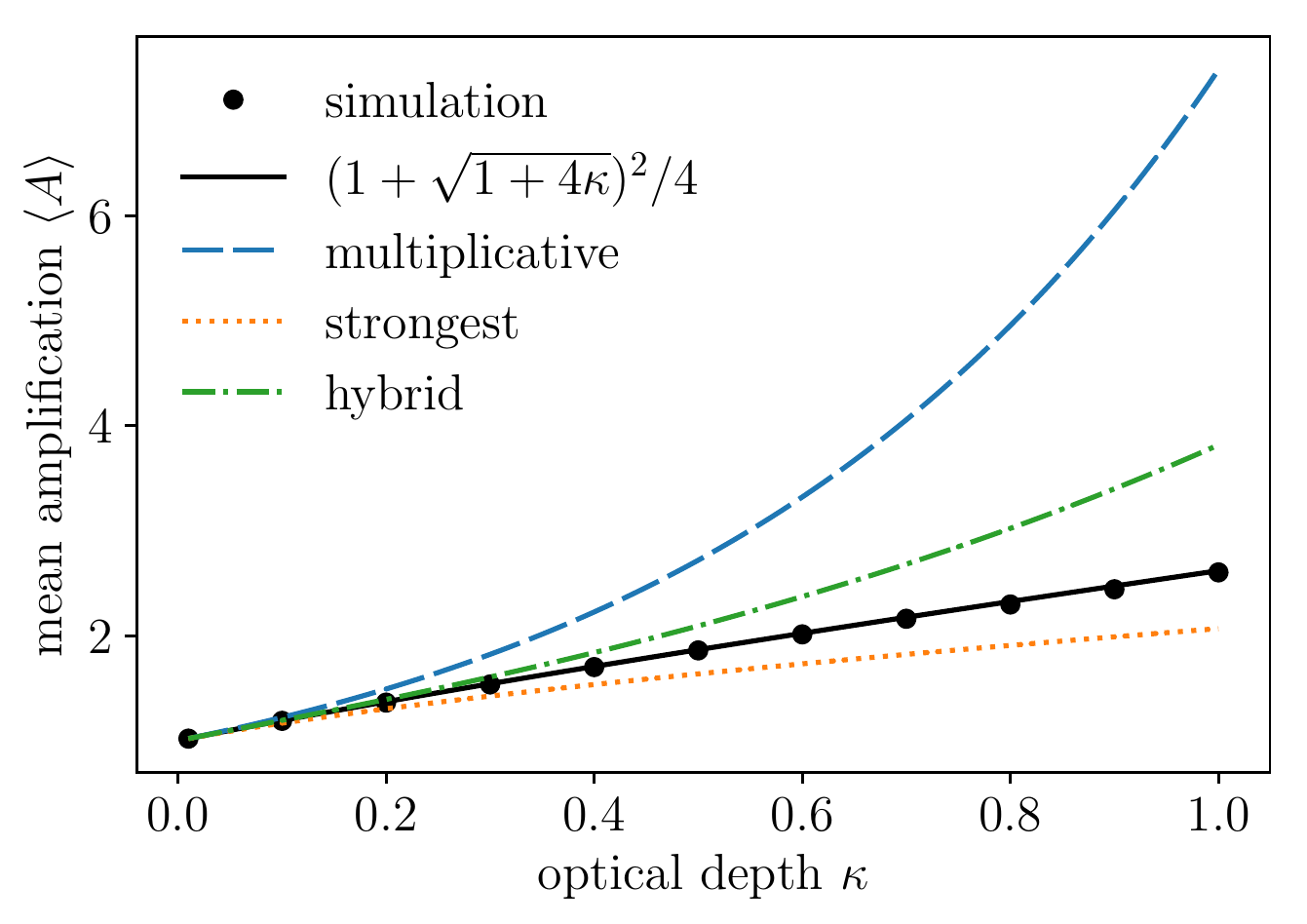}
\caption{Mean amplification~$\ev{A}$ as a function of optical depth~$\kappa$. Black disks indicate results from inverse ray shooting, while the various lines depict the models investigated in this article: multiplicative~$\ev{A}\e{m}$ (blue long-dashed); strongest-lens~$\ev{A}\e{s}$ (orange dotted); and hybrid~$\ev{A}\e{h}$ (green dot-dashed). The black solid line is the theoretical estimate~\eqref{eq:A_mean_est} of the simulated average.}
\label{fig:Amean}
\end{figure}

Figure~\ref{fig:Amean} shows the simulated mean amplification~$\ev{A}$ as a function of optical depth~$\kappa$, and compares it with the three models (strongest lens, multiplicative, and hybrid) considered in this article. In the low-optical depth regime, $\ev{A}\approx 1+2\kappa$, and the three models are in excellent agreement with the simulation. For larger optical depths, all three models fail. As expected, the strongest-lens approximation underestimates~$\ev{A}$ while the multiplicative approach overestimates it. The hybrid model lies in between, but it still overestimates $\ev{A}$. However, since the estimate~\eqref{eq:A_mean_est} provides an excellent fit to the simulation, the failure of $\ev{A}\e{s}, \ev{A}\e{m}, \ev{A}\e{h}$ to predict the correct $\ev{A}$ must be partly attributed to finite-size effects. Indeed, all three models assume an infinite number of lenses in an infinite Universe, whereas the numerical setup is finite.

\subsection{Comparing PDF models}
\label{subsec:comparing_PDF}

Let us now confront the three PDF models~$p\e{s}(A)$, $p\e{m}(A)$, $p\e{h}(A)$ to the simulations. For the sake of completeness, we add a fourth one, empirically proposed by Rauch in 1991~\cite{1991ApJ...374...83R}. For low optical depths, the following expression was found to provide a good fit to Monte-Carlo simulations, aiming to determine microlensing amplification statistics in an expanding Universe filled with point lenses,
\begin{equation}\label{eq:Rauch}
p\e{R}(A) \define 2\kappa\e{eff}  \pac{\frac{1-\ex{-b(A-1)}}{A^2-1}}^{3/2} \ .
\end{equation}
In Eq.~\eqref{eq:Rauch}, $\kappa\e{eff}$ and $b$ are two parameters fixed by the conditions that $p\e{R}$ is normalized to unity, and that $\ev{A}\e{R}$ gives the correct mean amplification (as given by the simulation). Rauch's fitting formula, combined with flux conservation, has been used in Refs.~\cite{Seljak:1999tm, Metcalf:1999qb, Metcalf:2006ms}, and even more recently in Ref.~\cite{Zumalacarregui:2017qqd} to set constraints on the abundance of primordial black holes from supernova lensing (see also Ref.~\cite{Garcia-Bellido:2017imq}).

\begin{figure}[h!]
\centering
\includegraphics[width=\columnwidth]{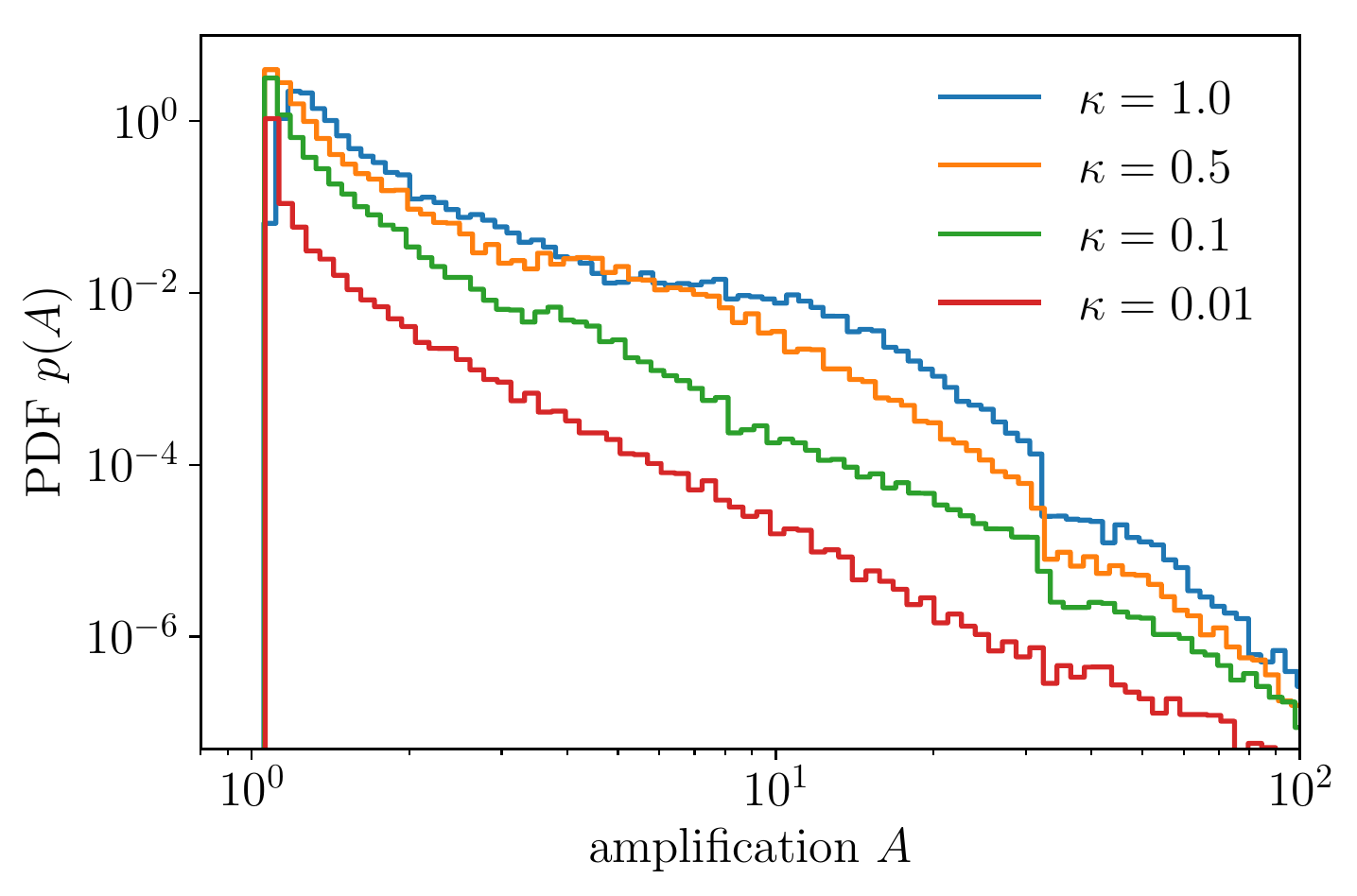}
\caption{Numerical amplification PDFs obtained for four different values of the optical depths~$\kappa$, from top to bottom: $1, 0.5, 0.1, 0.01$.}
\label{fig:numerical_PDFs}
\end{figure}

\begin{figure*}
\centering
\parbox{0.75\columnwidth}{
\flushleft{
\includegraphics[width=0.55\columnwidth]{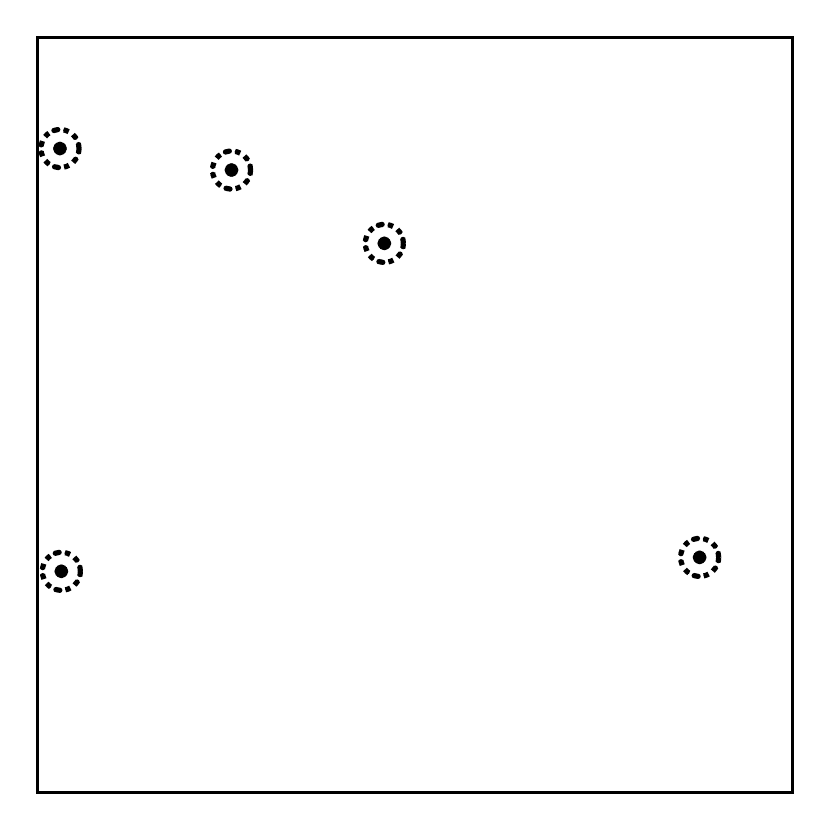}
} \\[-3mm]
\includegraphics[width=0.75\columnwidth]{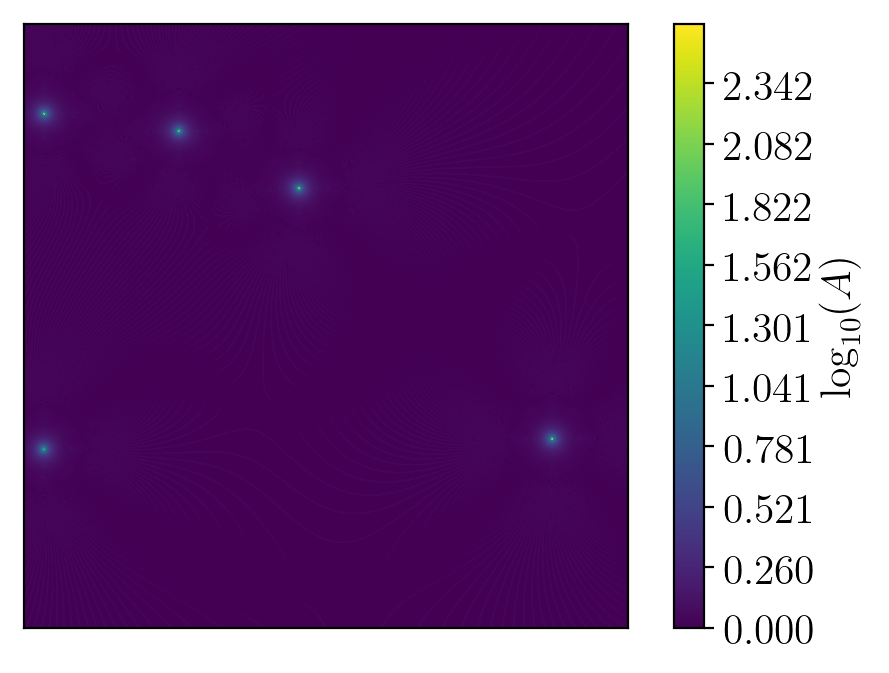}
% size ratio = 1.35
}
\hspace{1cm}
\parbox[c]{1.05\columnwidth}{
\includegraphics[width=1.05\columnwidth]{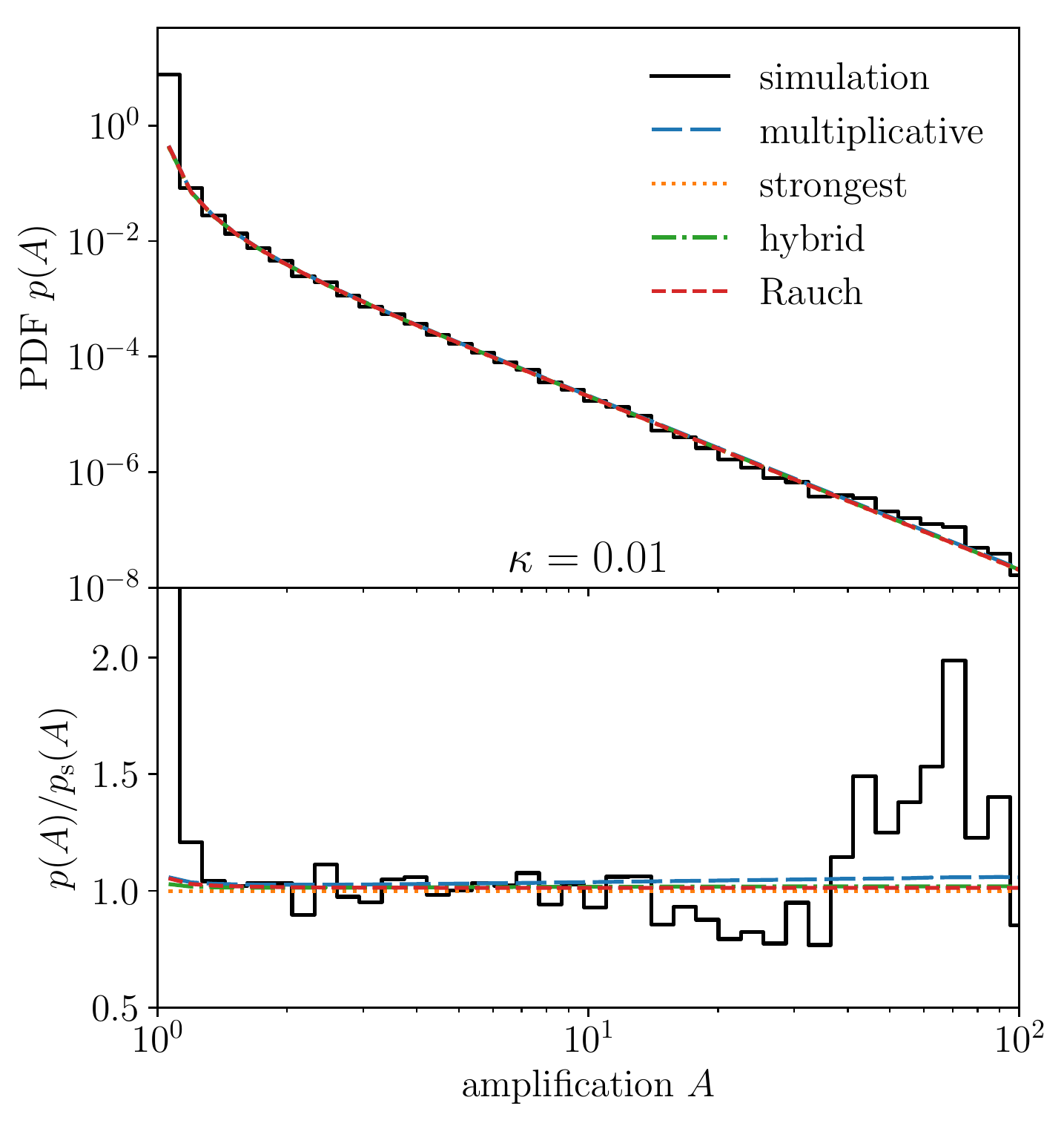}
}
\caption{Amplification map and PDF for the optical depth $\kappa=0.01$. \textit{Left panel}: position of the $N\e{l}=5$ lenses with their individual Einstein radii (dashed lines), and the corresponding amplification map in the source plane. \textit{Right panel}: PDF of the amplification. In the top panel are shown numerical results obtained from inverse ray shooting (black, solid), as well as four analytic models: multiplicative~$p\e{m}(A)$ (blue, long-dashed); strongest lens~$p\e{s}(A)$ (orange, dotted); hybrid~$p\e{h}(A)$ (green, dot-dashed); and Rauch's fitting function~$p\e{R}(A)$ (red, short-dashed). The bottom panel shows the relative difference $|p\e{mod}-p\e{sim}|/p\e{sim}$ between each analytic model and the simulation.}
\label{fig:kappa001}
\end{figure*}

\begin{figure*}
\centering
\parbox{0.75\columnwidth}{
\flushleft{
\includegraphics[width=0.55\columnwidth]{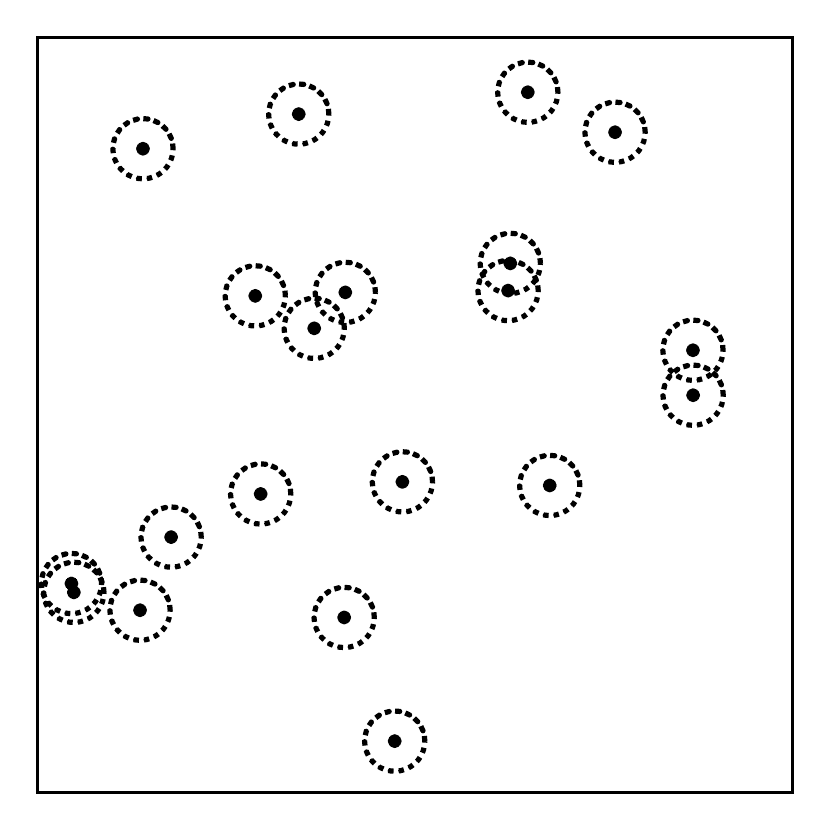}
} \\[-3mm]
\includegraphics[width=0.75\columnwidth]{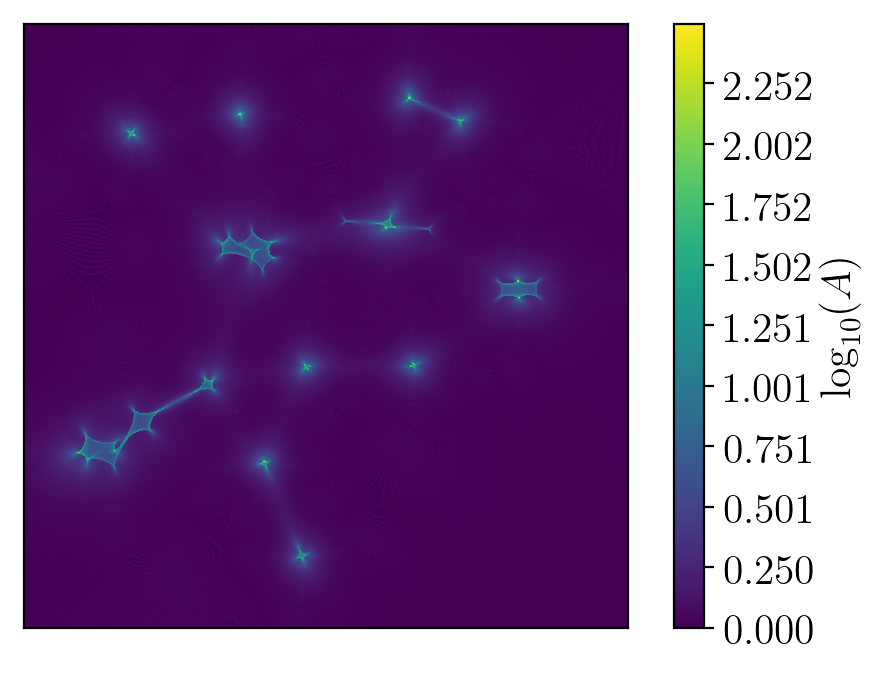}
% size ratio = 1.35
}
\hspace{1cm}
\parbox[c]{1.05\columnwidth}{
\includegraphics[width=1.05\columnwidth]{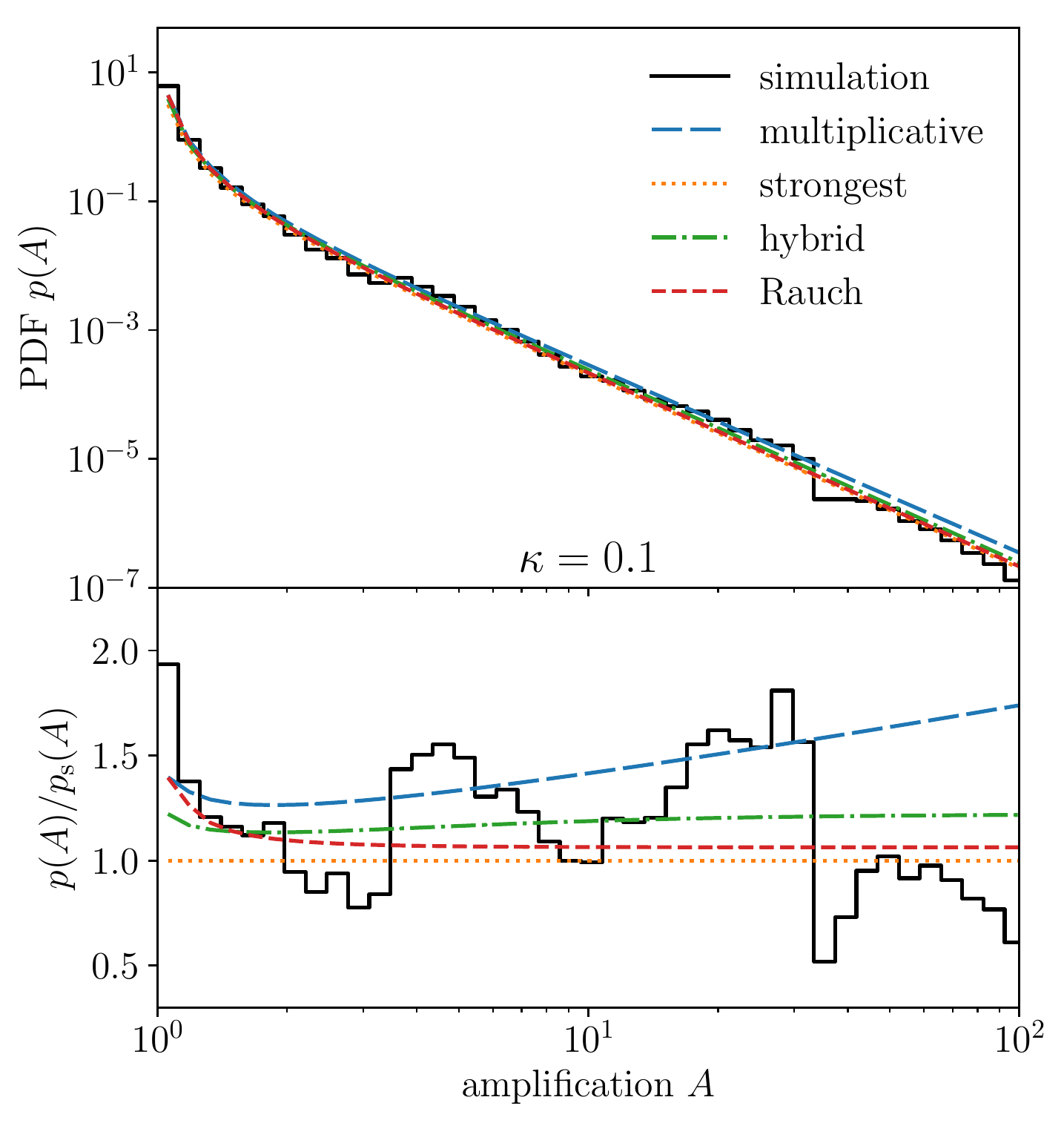}
}
\caption{Same as Fig.~\ref{fig:kappa001}, but with $N\e{l}=20$ lenses corresponding to a total optical depth~$\kappa=0.1$.}
\label{fig:kappa01}
\end{figure*}

\begin{figure*}
\centering
\parbox{0.75\columnwidth}{
\flushleft{
\includegraphics[width=0.55\columnwidth]{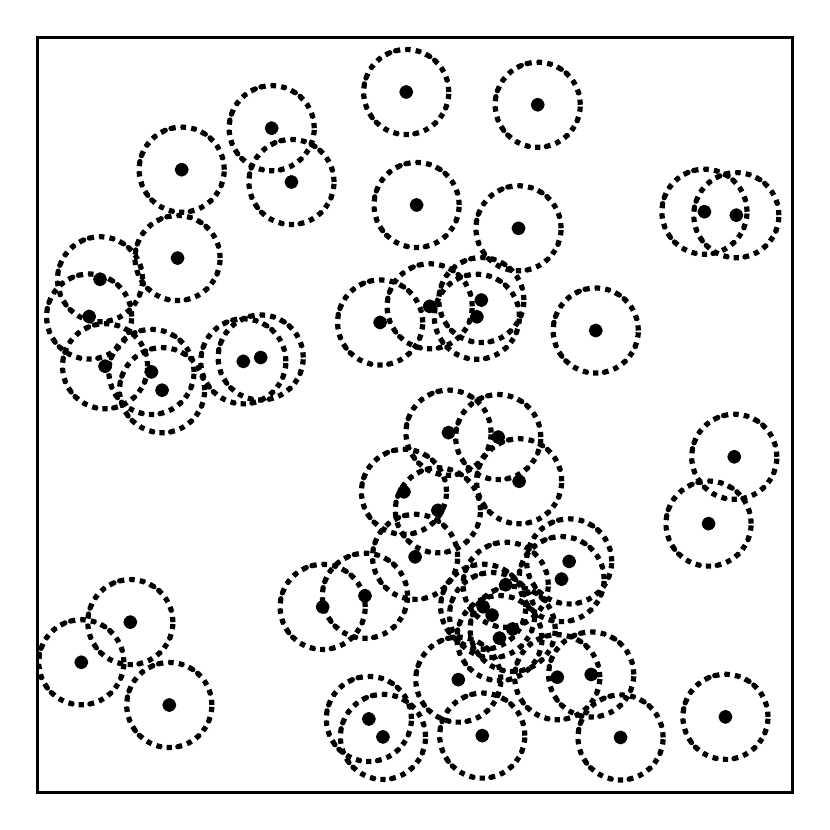}
} \\[-3mm]
\includegraphics[width=0.75\columnwidth]{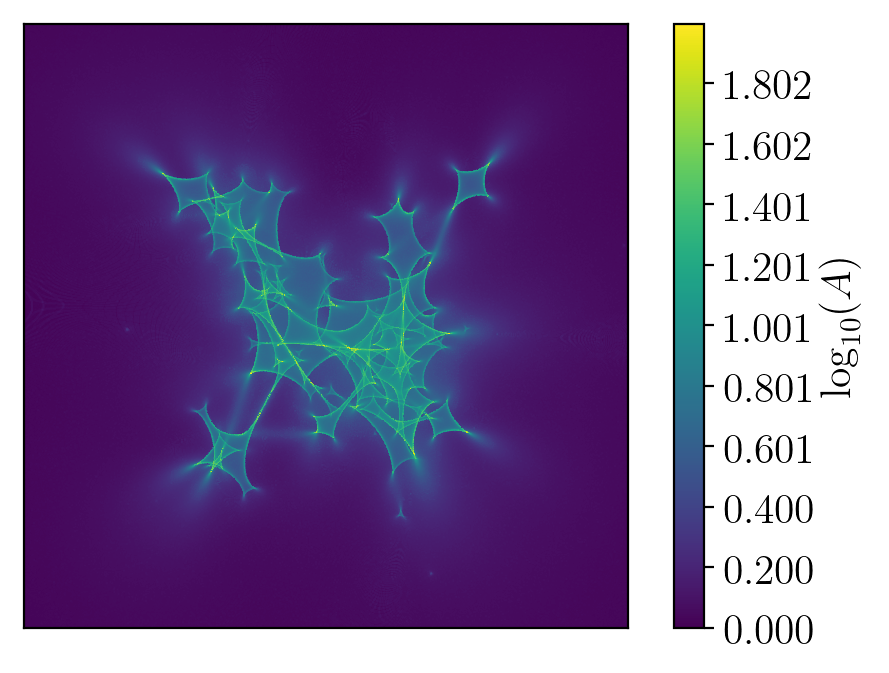}
% size ratio = 1.35
}
\hspace{1cm}
\parbox[c]{1.05\columnwidth}{
\includegraphics[width=1.05\columnwidth]{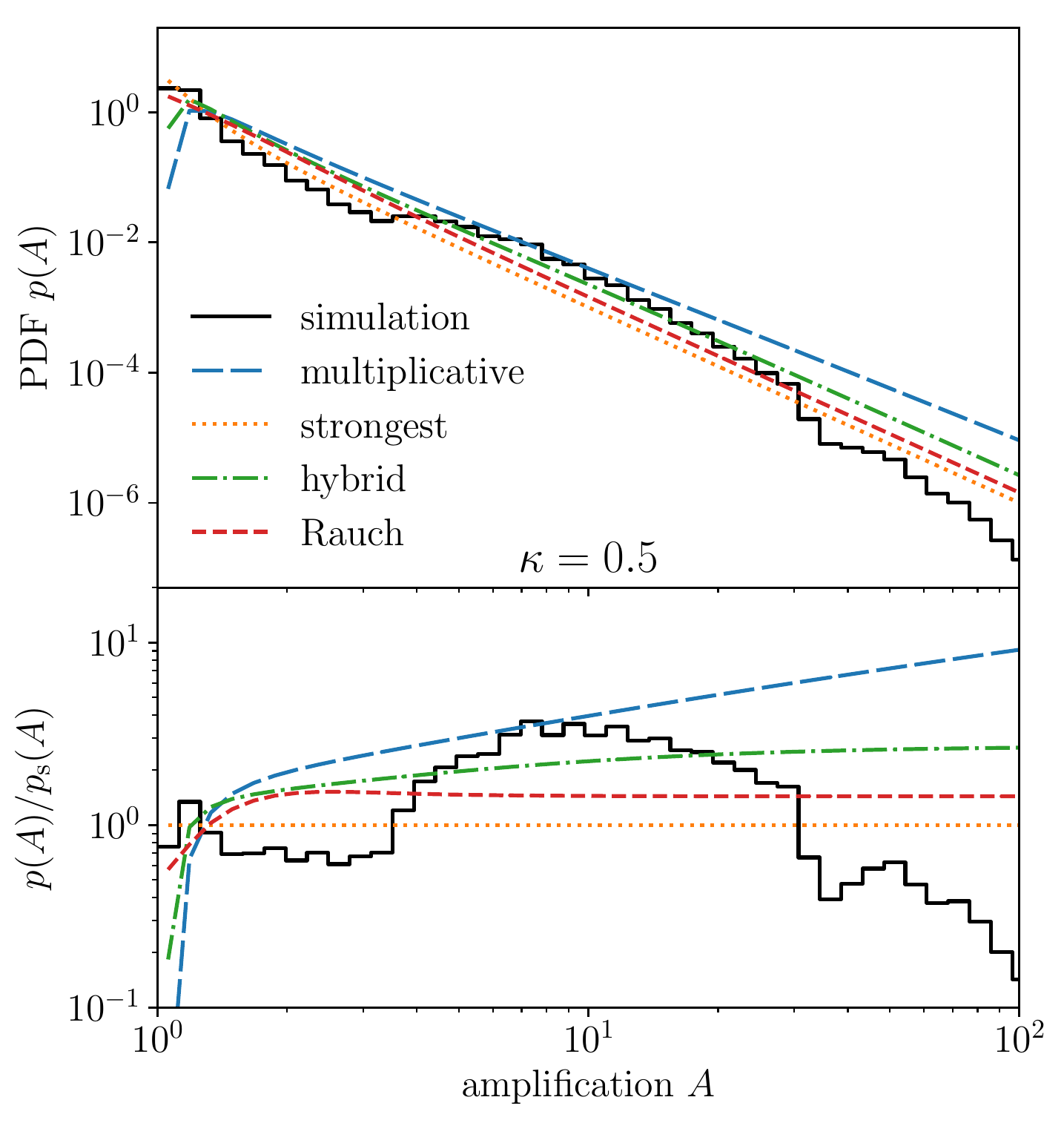}
}
\caption{Same as Fig.~\ref{fig:kappa001}, but with $N\e{l}=50$ lenses corresponding to a total optical depth~$\kappa=0.5$.}
\label{fig:kappa05}
\end{figure*}

\begin{figure*}
\centering
\parbox{0.75\columnwidth}{
\flushleft{
\includegraphics[width=0.55\columnwidth]{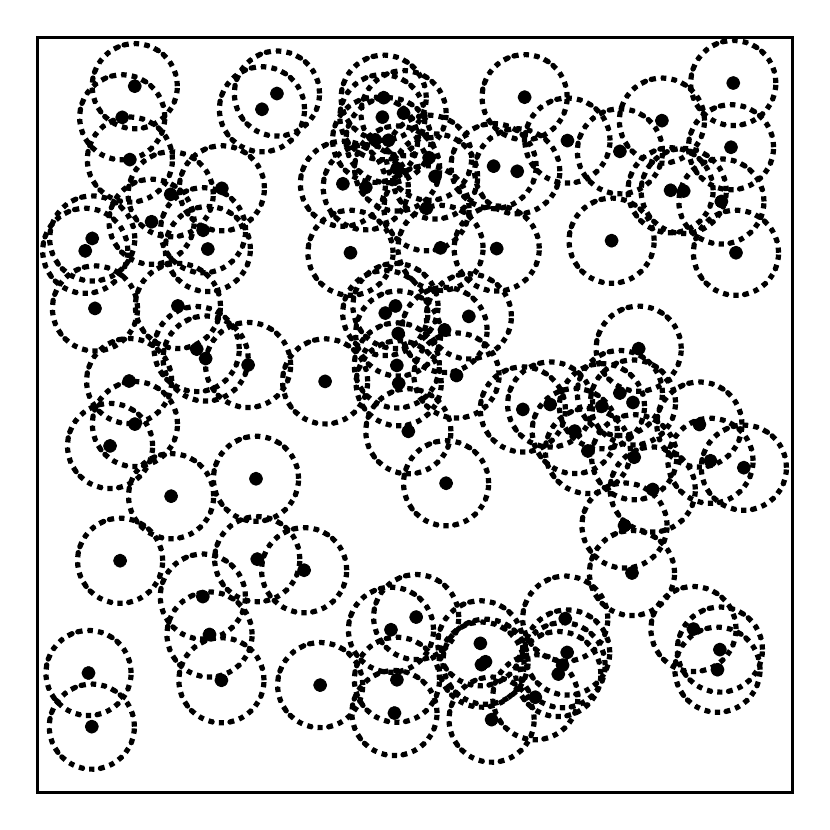}
} \\[-3mm]
\includegraphics[width=0.75\columnwidth]{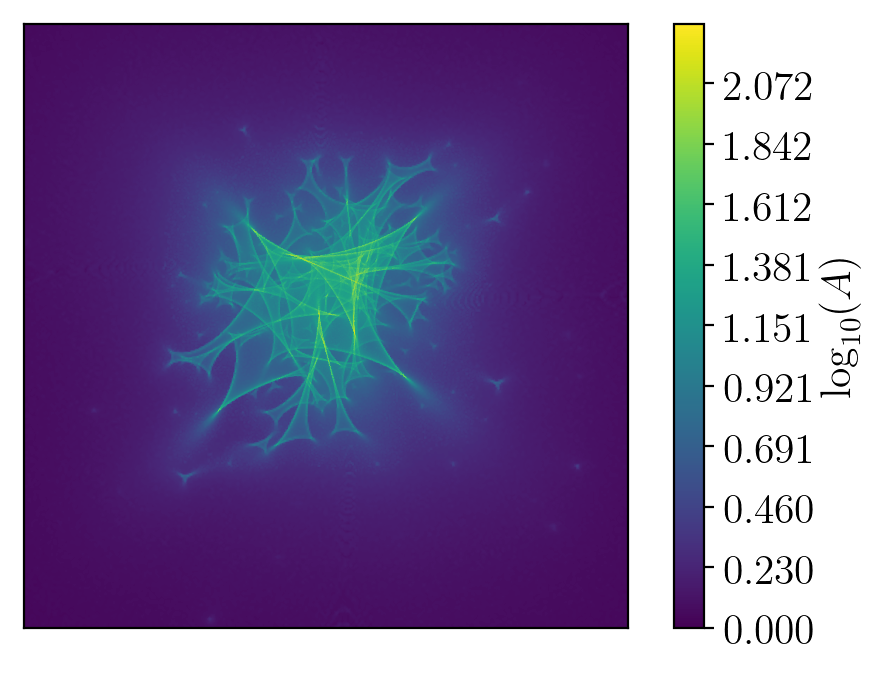}
% size ratio = 1.35
}
\hspace{1cm}
\parbox[c]{1.05\columnwidth}{
\includegraphics[width=1.05\columnwidth]{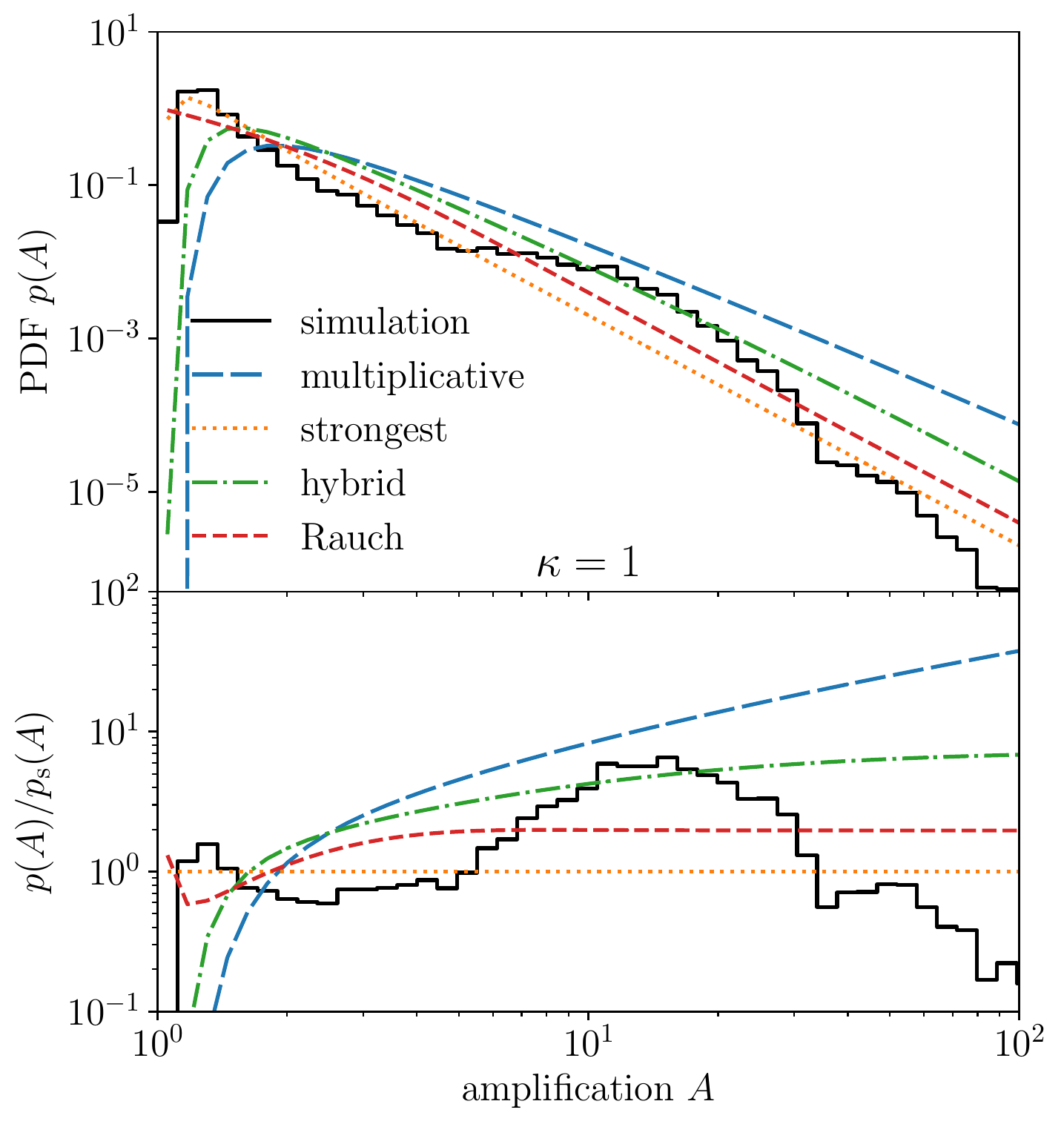}
}
\caption{Same as Fig.~\ref{fig:kappa001}, but with $N\e{l}=100$ lenses corresponding to a total optical depth~$\kappa=1$.}
\label{fig:kappa1}
\end{figure*}

We consider four values of the optical depth, $\kappa=0.01, 0.1, 0.5, 1$. For each value, the parameters of the simulation, Rauch's fitting function, and the mean amplification, are summarized in Table~\ref{tab:parameters}. The four numerical amplification PDFs are shown together in Fig.~\ref{fig:numerical_PDFs}, and individually compared to the analytic models in Figs.~\ref{fig:kappa001}, \ref{fig:kappa01}, \ref{fig:kappa05}, \ref{fig:kappa1}.

\begin{table}[h!]
\centering
\begin{tabular}{c|ccccc}
\hline
\hline
$\kappa$ & $N\e{l}$ & $N$ & $\ev{A}$ & $\kappa\e{eff}$ & $b$ \\ 
\hline
0.01 & 5 & $10^8$ & $1.019$ & $0.0101$ & $3340$ \\ 
%\hline 
0.1 & 20 & $10^8$ & $1.189$ & $0.106$ & $41.5$ \\ 
%\hline 
0.5 & 50 & $10^8$ & $1.859$ & $0.719$ & $2.53$ \\ 
%\hline 
1 & 100 & $2\times 10^7$ & $2.602$ & $1.96$ & $0.818$ \\ 
\hline
\hline
\end{tabular}
\caption{Parameters used for the simulations: optical depth~$\kappa$, number of lenses~$N\e{l}$, number of rays shot~$N$. We also indicate the mean amplification~$\ev{A}$, and the parameters~$\kappa\e{eff}, b$ of Rauch's fitting formula~\eqref{eq:Rauch}.}
\label{tab:parameters}
\end{table}

As expected, for low optical depths $(\kappa=0.01, 0.1)$, all four models~$p\e{s}, p\e{m}, p\e{h}, p\e{R}$ essentially coincide, and are in very good agreement with numerical results. However, in the optically-thick regime ($\kappa=0.5, 1$), all four models heavily fail to reproduce the actual behavior of $p(A)$. Specifically, the models tend to underestimate the probability of large amplifications ($A\sim 10$), and overestimate probability of very large amplifications ($A>100$).

The reason for this failure is nonlinear lens-lens coupling, which none of the four models really accounts for. The impact of lens-lens coupling is clearly visible on the amplification maps. As $\kappa$ increases, the map's aspect changes from a set of small isolated regions, where the amplification can become extremely large, to an intricate and cuspy caustic network, where large areas are characterized by intermediate amplifications, but where it is difficult to access very high values of $A$. As already observed in the literature~\cite{1984JApA....5..235N, 1987ApJ...319....9S, 1992ApJ...389...63M, 1997ApJ...489..508K}, this explains the leaking from very large to large amplifications, compared to what would be na\"{i}vely expected from approaches where the lenses are independent.

Of the four analytic models, the multiplicative approach is probably the worst, while the strongest-lens approach may be considered the least bad. However, in order to accurately model the amplification PDF in the optically-thick regime, it is necessary to properly tackle the problem of lens-lens coupling. Such a program, already initiated by other authors~\cite{1987ApJ...319....9S, 1997ApJ...489..508K, 1997ApJ...489..522L}, is beyond the scope of the present article, but shall be addressed in a future work.

\subsection{Extended sources}
\label{subsec:extended_sources}

Let us finally consider the impact of the finite size of sources. In Sec.~\ref{sec:extended_sources}, this effect was discussed for homogeneous disk-sources with angular radius~$\sigma$. In the present numerical setup, however, \emph{square sources} are more easily implemented. In order to connect the former theoretical results to the latter numerical ones, we assume that a disk source with angular area $\pi\sigma^2$ is mostly equivalent to a square source with the same area.

\begin{figure}[t]
\centering
\includegraphics[width=\columnwidth]{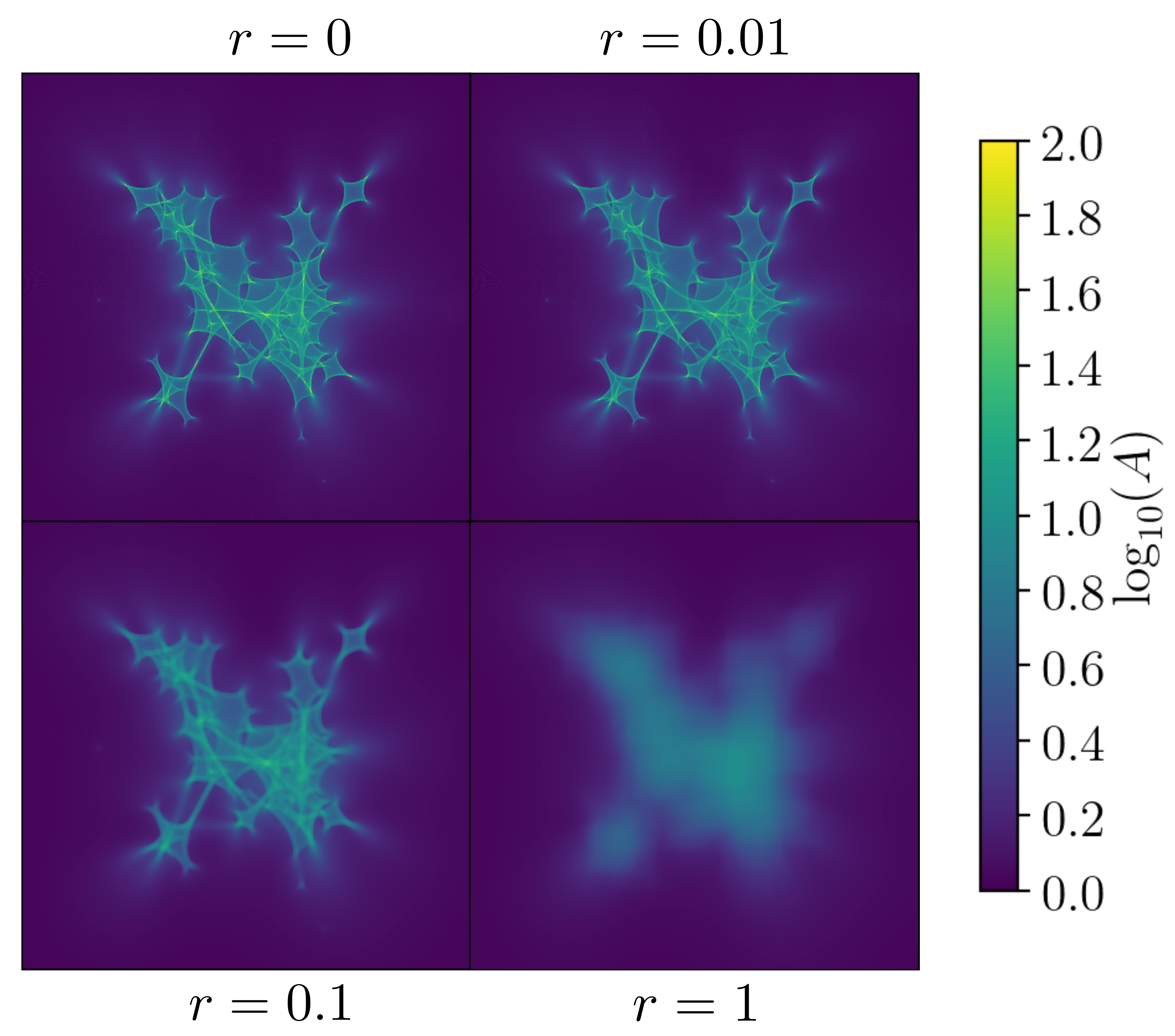}
\caption{Amplification map for the same distribution of lenses as
in Fig.~\ref{fig:kappa05} ($\kappa=0.5$) but for sources with various sizes, corresponding to the reduced radii $r=0,10^{-2},10^{-1}, 1$.}
\label{fig:maps_ES_kappa05}
\end{figure}

\begin{figure}[t]
\centering
\includegraphics[width=\columnwidth]{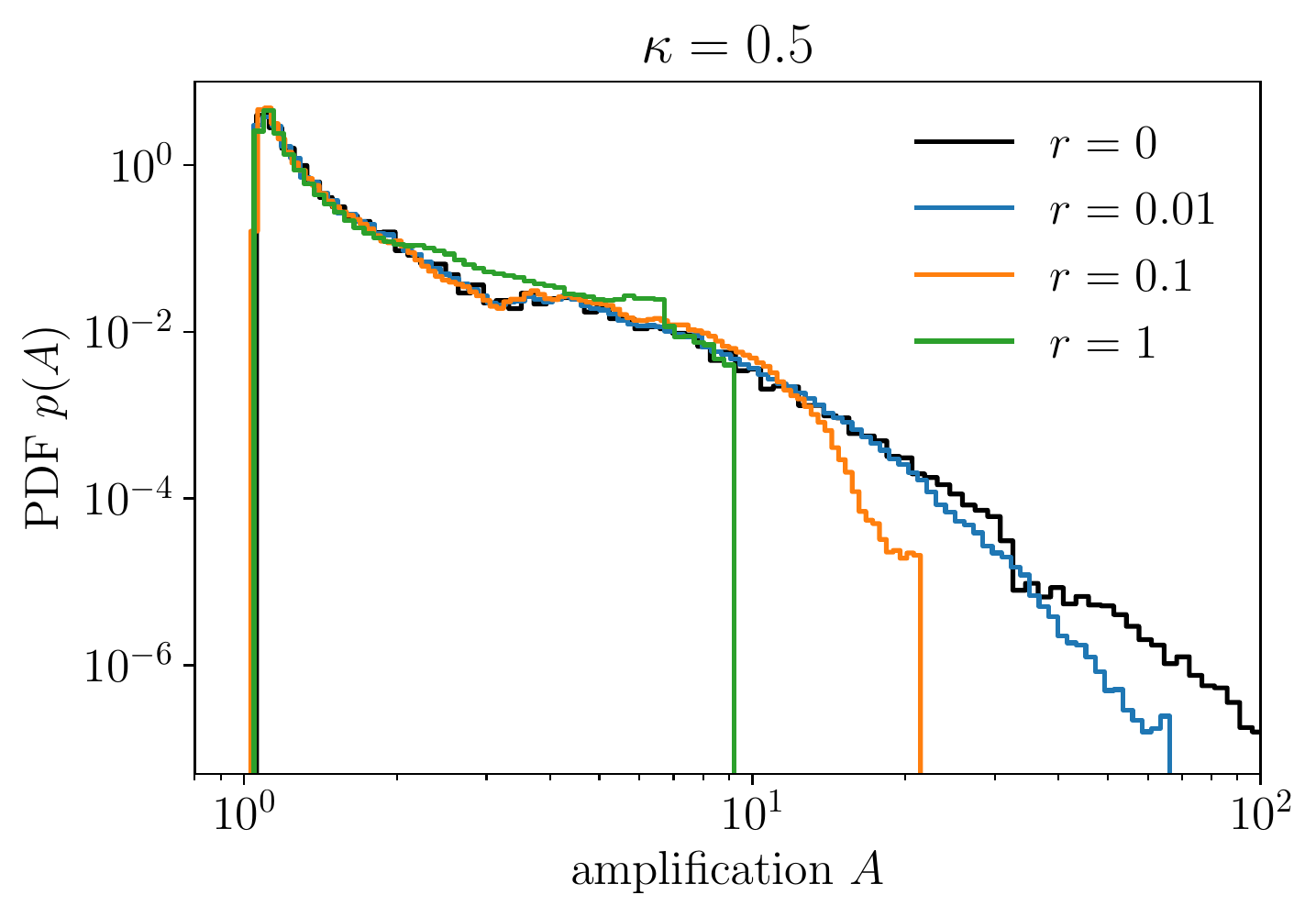}
\caption{Amplification PDFs corresponding to the maps of Fig.~\ref{fig:maps_ES_kappa05}.}
\label{fig:pdf_kappa05_ES}
\end{figure}

As mentioned in Sec.~\ref{sec:extended_sources}, the microlensing amplification map for extended sources is a smoothed version of the the map obtained with point sources. This property is clearly illustrated in Fig.~\ref{fig:maps_ES_kappa05} for $\kappa=0.5$. Four different source sizes are represented, $r=0, 10^{-2}, 10^{-1}, 1$, where $r=\sigma/\eps$ is the reduced size of the source. We are using square sources with edge $\ell=\sqrt{\pi}\sigma$, so that $\ell^2=\pi\sigma^2=\pi r^2\eps^2$. As the source's radius increases, the map gets smoother; in particular, the maximum amplification decreases. The impact of that smoothing on the amplification PDF is depicted in Fig.~\ref{fig:pdf_kappa05_ES}.

\begin{figure}[t!]
\centering
\includegraphics[width=\columnwidth]{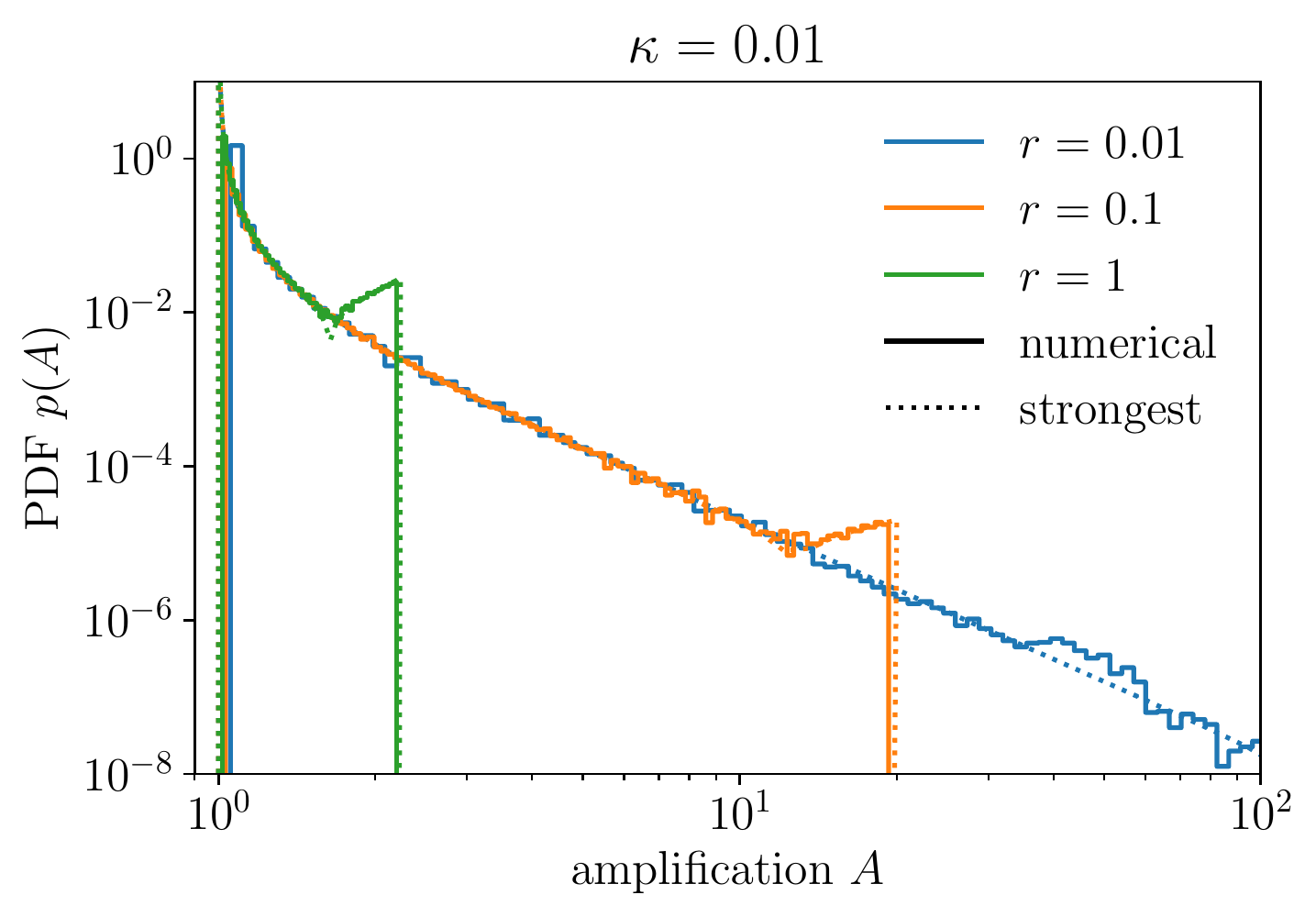}
\caption{Amplification PDFs for extended sources and low optical depth $\kappa=0.01$ (same setup as in Fig.~\ref{fig:kappa001}). The multiplicative model is comparable to the strongest-lens model in that regime.}
\label{fig:pdf_kappa001_ES}
\end{figure}

\begin{figure}[h!]
\centering
\includegraphics[width=\columnwidth]{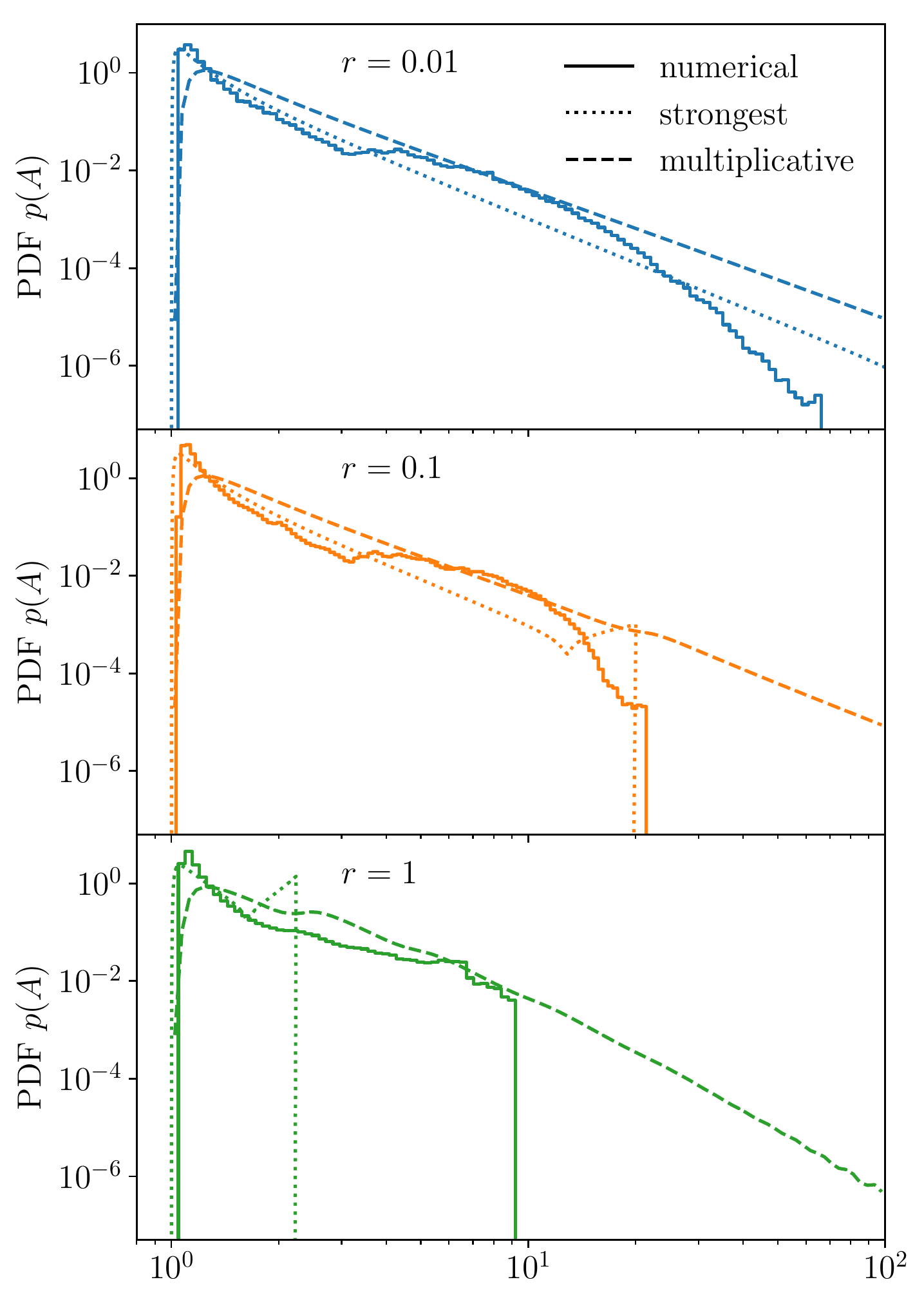}
\caption{Same as Fig.~\ref{fig:pdf_kappa05_ES}, but where numerical results (solid lines) are confronted to the predictions of the strongest-lens model (dotted) and of the multiplicative model (dashed). Just like in the point-source case, simple models largely fail at reproducing the amplification PDF for large optical depths.}
\label{fig:pdf_kappa05_ES_with_models}
\end{figure}

What is the performance of the strongest-lens and multiplicative  models to describe $p(A)$? For low optical depths, we know from Eq.~\eqref{eq:p_m=p_s_ES} that both models coincide up to terms of order $\kappa^2$. Their common prediction is in excellent agreement with the numerical results, as we can see in Fig.~\ref{fig:pdf_kappa001_ES} for $\kappa=10^{-2}$. However, for higher optical depths, both models fail just like in the point-source case, see Fig.~\ref{fig:pdf_kappa05_ES_with_models}. This is especially true for $r=1$, where the strongest-lens model underestimates the maximum amplification by almost a factor $5$. This is due to the fact that, in the strongest-lens model, the light beam can only enclose one lens, while for high optical depths the beam would typically enclose multiple lenses. The multiplicative model does not predict any maximum amplification, thereby missing an important property of the actual $p(A)$.

%%%%%%%%%%%
\section{Conclusion}
\label{sec:conclusion}
%%%%%%%%%%%

This article reviewed simple analytic models for the probability distribution of microlensing amplifications due to a set of microlenses. All the models are parameterized by a single quantity, namely the optical depth~$\kappa=\Sigma\pi\ev[1]{r\e{E}^2}$, which depends on the surface density of lenses~$\Sigma$ and their mean Einstein radius~$r\e{E}$. The optical depth quantifies the expected amount of lensing caused by the system. It also coincides with the weak-lensing convergence which would be observed if the lenses' mass was smoothly distributed.

In the strongest-lens model, the amplification~$A=A\e{s}$ is only due to the lens whose dimensionless impact parameter with respect to the line of sight is the smallest. In order to allow for the combined effect of many lenses, we considered a multiplicative model. In that approach, the net amplification is assumed to be the product of every individual lens amplification, $A=A_1 A_2 \ldots A_N$. Since the existing derivation of the amplification PDF in that context was incorrect, we proposed an alternative one, and found the same final result. Finally, we considered a hybrid model where the net amplification is the product of the strongest individual amplification with the mean principal amplification of the other lenses, $A=A\e{s}\bar{A}^+\e{w}$. While calculations were mostly conducted assuming point sources, we also derived the finite-source corrections to the strongest-lens and multiplicative models.

These three simple models were confronted to two-dimensional inverse-ray-shooting simulations. For low optical depths, $\kappa<0.1$, they essentially coincide and are in excellent agreement with numerical results. They also predict the correct behavior of the mean amplification, $\ev{A}=1+2\kappa+\mathcal{O}(\kappa^2)$. However, for large optical depths, all three models fail to reproduce both the mean amplification and the main features of its PDF. In particular, the probability of large amplifications ($A\sim 10$) is underestimated, while the probability of very large amplifications ($A>100$) is overestimated. While the misestimate of the mean amplification can easily be attributed to the finite size of the map, the mismatch of the full PDF comes from a deeper modeling limitation. Namely, none of the models properly account for lens-lens coupling, which cannot be neglected in the optically-thick regime.

An accurate description of the microlensing amplification PDF is essential for using this observable as a reliable test of the nature and small-scale distribution of dark matter. This is particularly true for large optical depths, where most of the signal is expected. In future work, we aim to efficiently model the impact of lens-lens coupling, in order to produce realistic amplification statistics in a clumpy Universe.

%%%%%%%%%%%%%%%%%%%%%%%%%%%%%%%%%%%%%%%%%%%%%%%%%%%
\acknowledgements

PF thanks Jonathan Blazek and David Harvey for discussions and encouragements. We also thank the anonymous referee of Physics of the Dark Universe, for their constructive comments that significantly enriched the content of this article. PF acknowledges the financial support of the Swiss National Science Foundation. PF received the support of a fellowship from ``la Caixa'' Foundation (ID 100010434). The fellowship code is LCF/BQ/PI19/11690018. JGB thanks the CERN TH-Division for hospitality during his sabbatical, when this project was initiated, and acknowledges support from the Research Project FPA2015-68048-03-3P [MINECO-FEDER] and the Centro de Excelencia Severo Ochoa Program SEV-2012-0597. He also acknowledges support from the Salvador de Madariaga Program, Ref. PRX17/00056.

%%%%%%%%%%%%%%%%%%%%%%%%%%%%%%%%%%%%%%%%%%%%%%%%%%%

\appendix
%%%%%%%%%%%%%%%%%%%%%%%%%%%%
\section{Pei's erroneous derivation of $p_{\rm m}(A)$}
\label{app:Pei}
%%%%%%%%%%%%%%%%%%%%%%%%%%%%

This appendix reproduces Pei's derivation of the multiplicative amplification PDF $p\e{m}(A)$, where some mistakes are corrected, and the remaining weaknesses are pointed out.

\subsection{Setup and notation}

Consider a source located at redshift $z\e{s}$, and randomly distributed lenses with redshifts $z\in[0,z\e{s}]$, mean number density~$n(z)$, and possibly different masses. We introduce the following notation:
\begin{itemize}
\item $p(A,z)\,\dd A$ is the probability that the lenses within $[0,z]$ yield a total amplification between $A$ and $A+\dd A$. Thus, $p(A)=p(A,z\e{s})$.
\item $q(A,\dd z)\,\dd A$ is the probability that the lenses within $[z, z+\dd z]$ yield a total amplification between $A$ and $A+\dd A$.
\item $f(A,z)\,\dd A\,\dd z$ is the number of lenses\footnote{This quantity corresponds to $\rho(A,z|z\e{s})$ in the notations of \Pei.} between $z$ and $z+\dd z$ generating an individual amplification between $A$ and $A+\dd A$.
\end{itemize}

It is tempting to identify $q$ with $\partial p/\partial z$. However, this cannot be true, since $q$ is a probability density, hence normalized to 1, while the integral of $\partial p/\partial z$ over $A$ vanishes.

\subsection{Integrodifferential equation for $p(A,z)$}

The actual relationship between $p$ and $q$ comes from the prescription~\eqref{eq:combining_p_A} for the multiplicative combination between two groups of lenses. Namely, splitting the lenses of $[0,z+\dd z]$ between $[0,z]$ and $[z,z+\dd z]$, we find
\begin{equation}\label{eq:Pei_start}
p(A, z+\dd z) = \int_1^A \frac{\dd A'}{A'} \; q(A',\dd z) \, p(A/A',z) \ .
\end{equation}
The tricky part then consists in relating $q$ to $f$. One can arguably decompose $q$ as follows,
\begin{equation}
q(A,\dd z)  = \sum_{k=0}^\infty q_k(A, \dd z) \ ,
\end{equation}
where $q_k(A, \dd z)$ is the probability that exactly $k$ lenses within $[z, z+\dd z]$ significantly contribute to the amplification $A$. As $\dd z\rightarrow 0$, it is rather intuitive that $q_k = \mathcal{O}(\dd z^k)$; thus, at first order in $\dd z$, we must have
\begin{align}
q(A, \dd z)
&= q_0(A, \dd z) + q_1(A, \dd z) + \mathcal{O}(\dd z^2) \\
\label{eq:expansion_q}
&= C \delta(A-1) + f(A, z) \dd z + \mathcal{O}(\dd z^2) \ ,
\end{align}
where $C$ is a constant to be determined. Indeed, if no lens contributes ($k=0$), then $A=0$, so that $q_0(A, \dd z)\propto \delta(A-1)$. The identification between $q_1$ and $f \dd z$ then comes from the definition\footnote{\Pei erroneously considers $q_1(A,\dd z)=\rho(A,z|z+\dd z)$} of $f$. We determine $C$ using the normalization of $q$,
\begin{equation}
C = 1 - \bar{f}(z)\,\dd z + \mathcal{O}(\dd z^2) \ ,
\quad
\bar{f}(z) \define \int_1^\infty \dd A \; f(A,z) \ .
\end{equation}
Note that, since $\bar{f}$ represents the total number of lenses per unit redshift, this quantity is generally \emph{infinite}! However, its combination with $\delta(A-1)$ and $f(A,z)$ should remain finite. Substituting the expression of $q$ in Eq.~\eqref{eq:Pei_start}, we get the following integrodifferential equation:\footnote{Equation~\eqref{eq:Pei_integro_differential}
 is corrected with respect to Eq.~(4) of \Pei, which includes confusions between $z$ and $z\e{s}$, as well as a mathematical mistake about derivatives of dependent integrals.}
\begin{equation}\label{eq:Pei_integro_differential}
\pd{p}{z} = \int_1^A \frac{\dd A'}{A'} \; f(A',z) \, p(A/A',z) - \bar{f}(z) p(A,z) \ .
\end{equation}
Alternatively, one can use logarithmic probabilities, $P(L)=A p(A)$, $F(L)=A f(A)$, with $L=\ln A$, to get
\begin{equation}\label{eq:Pei_integro_differential_log}
\pd{P}{z} = \int_0^L \dd L' \;  F(L',z)\, P(L-L',z) - \bar{f}(z) P(L,z)\ .
\end{equation}

\subsection{Solving the equation}

The main difficulty of Eq.~\eqref{eq:Pei_integro_differential_log} is the convolution product on its right-hand side. This is greatly simplified in Fourier space,
\begin{equation}\label{eq:Pei_diff_log}
\pd{\tilde{P}}{z} = \pac{\tilde{F}(K,z)- \tilde{F}(0,z)} \tilde{P}(K,z) \ ,
\end{equation}
where we used that
\begin{equation}
\tilde{F}(0,z)
= \int_{-\infty}^{\infty} \dd L \; F(L,z)
= \int_{0}^{\infty} \dd A \; f(A,z)
= \bar{f}(z) \ .
\end{equation}
Equation~\eqref{eq:Pei_diff_log} is easily integrated, and the result translated in terms of $p$, yielding
\begin{equation}\label{eq:Pei_result_1}
p(A) = \int_{-\infty}^{\infty} \frac{\dd K}{2\pi} \; A^{\i K-1} \exp \int_0^{z\e{s}} \dd z \pac{\tilde{F}(K,z)- \tilde{F}(0,z) } ,
\end{equation}
which is Eq.~(8) of \Pei.

\subsection{Determining $\tilde{F}(K,z)$}

The last step of the derivation consists in explicitly computing $\tilde{F}(K,z)$. For that purpose, we first determine $f(A,z)$, which by definition reads
\begin{equation}
f(A,z) = \frac{\dd^2 N}{\dd A \dd z} \ .
\end{equation}
For point lenses, we have seen in Sec.~\ref{subsec:point_lens} that the amplification only depends on the reduced impact parameter $u=b/r\e{E}$. Hence, for one lens, the region of the plane $z=\cst$ such that its amplification lies in $[A, A+\dd A]$, has an area
\begin{equation}
\dd \sigma_A =
2\pi b \dd b = \pi r\e{E}^2(z) \dd u^2 = \pi r\e{E}^2(z) \abs{\ddf{u^2}{A}} \dd A \ .
\end{equation}
The quantity $\dd \sigma_A/\dd A$ must be understood as a differential cross section. Besides, from Eq.~\eqref{eq:u_A}, one finds
\begin{equation}
\abs{\ddf{u^2}{A}} = \frac{2}{(A^2-1)^{3/2}} \ .
\end{equation}

Suppose that an infinity of such lenses are randomly distributed within $[z, z+\dd z]$, with density $\dd\Sigma/\dd z=\dd^3 N/\dd^2 S \dd z$. The average number of lenses within $\dd z$ causing an amplification within $\dd A$ is $\dd^2 N = \dd\Sigma \, \dd \sigma_A$,
%
%\begin{equation}
%\dd N = \Sigma \, \dd z \, \dd \sigma_A = \frac{2\pi r\e{E}^2 \Sigma}{(A^2-1)^{3/2}} \, \dd A \, \dd z \ ,
%\end{equation}
%
whence
\begin{equation}\label{eq:f_A}
f(A,z) = \frac{2\pi r\e{E}^2(z)}{(A^2-1)^{3/2}} \, \ddf{\Sigma}{z} \ .
\end{equation}
Here, we have implicitly assumed that all the lenses have the same Einstein radius~$r\e{E}$, i.e. the same mass. However, it is straightforward to generalize the above rationale for lenses with a spectrum of masses. The idea consists in replacing $\dd\Sigma/\dd z$ with $\dd^2\Sigma/\dd z\dd m$, the density of lenses per unit mass, and then integrating over $m$.

The Fourier transform
\begin{align}
\tilde{F}(K,z)
&= \int_{-\infty}^\infty \dd L \; \ex{-\i K L} F(L,z) \\
&= \int_1^\infty \dd A \; A^{-\i K} f(A,z)
\end{align}
does not exist, because the integral does not converge at the limit $A=1$. However, the combination $\tilde{F}(K,z)-\tilde{F}(0,z)$, as it appears in Eq.~\eqref{eq:Pei_result_1} does exist. Precisely,
\begin{equation}\label{eq:Pei_Fourier}
\int_1^\infty \dd A \; \frac{A^{-\i K}-1}{(A^2-1)^{3/2}}
= 1 - \sqrt{\pi}\, \frac{\Gamma(1+\i K/2)}{\Gamma(1/2+\i K/2)} \ .
\end{equation}
Therefore, the final result is
\begin{equation}\label{eq:Pei_result_final}
p(A) =
\ex{2\kappa}
\int_{-\infty}^{\infty} \frac{\dd K}{2\pi} \; A^{\i K-1}
\exp\pac{ -2\kappa \sqrt{\pi}\, \frac{\Gamma(1+\i K/2)}{\Gamma(1/2+\i K/2)} } ,
\end{equation}
with
\begin{equation}\label{eq:optical_depth_Pei}
\kappa = \int_0^{z\e{s}} \dd z \int_0^\infty \dd m \; \frac{\dd^2\Sigma}{\dd z \dd m} \, \pi r\e{E}^2 \ .
\end{equation}
Equation~\eqref{eq:Pei_result_final} is the main result of \Pei. It is remarkable that it perfectly matches the outcome of Sec.~\ref{subsec:derivation_mutiplicative}, despite the several issues of the original derivation.

%%%%%%%%%%%%%%%%%%%%%%%%
\bibliography{bibliography_microlensing.bib}
%%%%%%%%%%%%%%%%%%%%%%%%
\end{document}